\documentclass[12pt,aip,jcp,amsfonts,amsmath,amssymb,preprint]{revtex4-1}
\usepackage{amsfonts, amsmath, amssymb, amsthm}

\newcommand{\dd}{{\rm d}}
\newcommand{\e}[1]{\,{\rm e}^{#1}\,}
\newcommand{\ii}{{\rm i}}

\newcommand{\bs}[1]{{\boldsymbol #1}}
\newcommand{\xn}{\bs{x}_{\rm n}}
\newcommand{\xe}{\bs{x}_{\rm e}}
\newcommand{\mn}{m_{\rm n}}
\newcommand{\me}{m_{\rm e}}

\def\sign{{\operatorname{sign}}}

\usepackage{graphicx}

\graphicspath{{./Figures/}}

\begin{document}

% Use the \preprint command to place your local institutional report number
% on the title page in preprint mode.
% Multiple \preprint commands are allowed.
%\preprint{}

\title{Multiple Superadiabatic Transitions and Landau-Zener Formulas} %Title of paper

% repeat the \author .. \affiliation  etc. as needed
% \email, \thanks, \homepage, \altaffiliation all apply to the current author.
% Explanatory text should go in the []'s,
% actual e-mail address or url should go in the {}'s for \email and \homepage.
% Please use the appropriate macro for the type of information

% \affiliation command applies to all authors since the last \affiliation command.
% The \affiliation command should follow the other information.

\author{B D Goddard}
\email[]{b.goddard@ed.ac.uk}
\affiliation{School of Mathematics and Maxwell Institute for Mathematical Sciences,
University of Edinburgh, EH9 3FD, UK}

\author{T Hurst}
\affiliation{School of Mathematics and Maxwell Institute for Mathematical Sciences,
University of Edinburgh, EH9 3FD, UK}

% Collaboration name, if desired (requires use of superscriptaddress option in
%\documentclass).
% \noaffiliation is required (may also be used with the \author command).
%\collaboration{}
%\noaffiliation

\date{\today}

\begin{abstract}
We consider nonadiabatic systems in which the classical Born-Oppenheimer approximation
breaks down.  We present a general theory that accurately captures the full transmitted wavepacket
after multiple transitions through either a single or distinct avoided crossings, including phase
information and associated interference effects.  
Under suitable approximations, we recover both the celebrated Landau-Zener formula and standard surface-hopping 
algorithms.
Our algorithm shows excellent agreement with the full quantum dynamics for
a range of avoided crossing systems, and can also be applied to single full crossings with similar accuracy.
\end{abstract}


\maketitle

\section{Introduction}

The Born-Oppenheimer approximation (BOA)~\cite{BornOppenheimer27} is one of the most widely used methods used to study the quantum dynamics
of molecules.  Intuitively, it is motivated by the fact that the electrons are much lighter, and therefore much faster, than the nuclei,
and hence rapidly adjust their positions with respect to those of the nuclei.  This scale separation allows, in many cases, for the 
electronic and nuclear dynamics to be decoupled.  In particular, if the electrons start in a certain bound state, for a fixed
set of nuclei positions, then they should remain in this bound state even though the nuclei are slowly moving.  Hence the nuclear
dynamics can be determined by considering their motion on only one (electronic) potential energy surface. 

However, there are interesting situations in which the BOA breaks 
down~\cite{Domcke04,Domcke11,Nakamura12,Tully12}.  For example, in many photochemical processes the
nuclear motion cannot be restricted to a single potential energy surface because, for some nuclear configurations, two such 
surfaces become close, or even cross.  In the former case, known as an avoided crossing, 
the BOA is still valid to leading order (in the small parameter $\epsilon$, which is the square root of the ratio of the electronic
and nuclear masses), but the remaining corrections
are of fundamental interest and, in fact, determine the associated chemistry.  In the latter case, which generally takes
the form of conical intersections, the BOA breaks down completely.

Here we are primarily interested in cases where the transmitted wavepacket is (exponentially) small
~\cite{HagedornJoye01,HagedornJoye05,MartinezSordoni02}, for example when there is
an avoided crossing, or when the wavepacket does not pass directly over the conical intersection.  Such regimes are,
in some sense, generic, as avoided crossings are generic in 1D~\cite{VonNeumanWigner29}, 
and in higher dimensions the probability of an arbitrary
wavepacket exactly hitting a conical intersection is vanishingly small~\cite{Tully12}.  
In particular, we consider cases where the wavepacket
passes through multiple avoided crossings, or repeatedly through the same crossing.  In such cases the transmitted wavepackets
can interfere, and thus it is necessary to understand their phases.  This suggests that a full quantum mechanical
treatment of the problem is required.  However, in even moderate dimensions, such treatments are numerically intractable,
especially for multiple, coupled electronic potential surfaces.

In order to overcome this, a range of coupled quantum-classical and semiclassical methods have been developed.
These include the 
multiple-spawning wavepacket method~\cite{BenNunMartinez98,BenNunQuennevilleMartinez00,VirshupChenMartinez12},
the frozen Gaussian wavepacket method~\cite{Heller91}, 
Ehrenfest dynamics~\cite{Mclachlan64,MeyerMiller79,SawadaNitzanMetiu85},
and the semiclassical initial value representation~\cite{Miller70,Kreek74,Miller01}.
The main advantage of such schemes is the significantly reduced computational cost.  The main disadvantage, at least with
respect to the problem at hand, is the lack of phase information from almost all such schemes.
Along with those mentioned above, one of the most widely-used quantum-classical approaches is surface hopping
~\cite{Tully71,MillerGeorge72,Stine76,Kuntz79,Tully90,HammesSchifferTully94,MullerStock97,FabianoGroenhofThiel08,
FermanianKammererLasser08,LasserSwart08,BelyaevLasserTriglia14,BelyaevDomckeLasserTriglia15},
in which particles are evolved 
under classical dynamics on a single surface
and can `hop' to other surfaces with a specified probability.  Perhaps the most common approach is to only allow hops
at points in the trajectory where the gap between energy surfaces has a local minimum (i.e.\ at an avoided crossing), and
the probability of the hop is given by a Landau-Zener (LZ) formula~\cite{Zener32,Landau65}.  Such methods give good results for a single transition, 
especially when the transmitted wavepacket is reasonably large, but fail completely when multiple transitions are involved,
due to the complete lack of phase information~\cite{FermanianKammererLasser17}.  We note here that there is at
least one such scheme~\cite{ChaiJinLiMorandi15} that does aim to retain the phase information, but this is limited to small gaps between the
potential energy surfaces, which in turn leads to large transmitted wavepackets.  The same restriction is true for other
mathematical approaches that lead to explicit formulae for the transmitted wavepacket; see e.g.\ Ref.~\cite{HagedornJoye05}.  It has
been shown that, if the gap scales with $\epsilon$, then the transitions are of order one and dominated by the Landau-Zener
factor~\cite{Hagedorn94,HagedornJoye98}.

An alternative approach, inspired by the work of Berry on superadiabatic representations~\cite{Berry90,BerryLim93},
considers the full quantum mechanical wavepacket.  These results, which are restricted to the semiclassical regime
where the nuclei move classically, were later made  rigorous~\cite{HagedornJoye04,BetzTeufel05-1,BetzTeufel05-1}.
It was later shown that, through the use of such superadiabatic
representations (which are generalisations of the well-known adiabatic representation), it is possible 
to derive a formula for the
transmitted wavepacket, including phase, at an avoided crossing
~\cite{BetzGoddardTeufeul09,BetzGoddard09,BetzGoddard11,BetzGoddardManthe16}.  
The associated algorithm requires only the quantum evolution of wavepackets 
on single energy surfaces.  Whilst this is still computationally demanding if one wants to solve the full Schr\"odinger equation,
there are approximate methods, such as Hagedorn wavepackets~\cite{Hagedorn81,Hagedorn94,Lubich08} or standard quantum
chemistry techniques such as MCTDH~\cite{MCTDHBook}, 
which make small relative errors and are
computationally much more tractable.  The associated algorithm has so far been applied to single transitions through avoided crossings~\cite{BetzGoddardTeufeul09,BetzGoddard09,BetzGoddard11}, and to multiple 
transitions of a single crossing in the case of the photodissociation of NaI~\cite{BetzGoddardManthe16}.  
The main goals here are to extend the methodology to multiple transitions through different avoided crossings and to 
systematically study the effects of making various approximations that lead to a LZ-like transition probability.  We will also
demonstrate that, although not designed to tackle such problems, the methodology can be successfully applied to single
transitions of full crossings.

We present an algorithm that has a number of advantages.  We have already mentioned:
(i) Preservation of phase information, which allows the accurate study of interference effects; 
(ii) Only evolution on a single surface is required, which significantly reduces the computational cost when compared to 
a fully-coupled system, whilst also allowing the use of state-of-the art numerical schemes. 
The main other benefits are:
(iii) Only the adiabatic surfaces (which are the most commonly obtained surfaces from quantum chemistry calculations)
are required, in particular there is no need for a diabatization scheme, or the determination of the adiabatic coupling 
elements; 
(iv) Such surfaces are only required locally, and thus can be computed on-the-fly; 
(v) The transmitted wavepacket is created instantaneously, and hence there is no reliance on complicated numerical 
cancellations of highly-oscillatory wavepackets, which are generally present in the adiabatic representation;
(vi) The methodology is easily extended to multiple adiabatic surfaces;
(vii) The derived formula is accurate for a wide range of potential energy gaps and small parameters $\epsilon$,
and for any semiclassical wavepacket, i.e.\ one of typical width or order $\sqrt{\epsilon}$.

There are, of course, also some disadvantages when compared to the more widely-used schemes:
(i) In order to capture the phase information, the one-level dynamics must retain at least some of their quantum nature,
and this is inherently more computationally demanding than the analogous classical dynamics;
(ii) In the full formalism, it is necessary to be able to extend the potential surface into the complex plane, at least
in the region of an avoided crossing.  This is essential to be able to accurately compute the transition probabilities.
However, in some regimes, for example when the LZ formula is accurate, we can bypass this requirement;
(iii) The scheme is, in principle, restricted to wavepackets that are semiclassical near the avoided crossing.  However,
due to the linearity of the Schr\"odinger equation, and as demonstrated in~\cite{BetzGoddardManthe16}, it is possible to `slice' the wavepacket
at the crossing.  However, this may be more problematic in higher dimensions;
(iv) As it stands, the method is restricted to 1D.  However, we have successfully extended it to higher dimensions through
 a slicing procedure [[REF 2D PAPER]].

To outline our approach,
we will first review the standard model for nonadiabatic transitions (Section~\ref{S:model}) 
and avoided crossings (Section~\ref{S:avoidedCrossings}).  
We will then, in Section~\ref{S:existing}, give
a brief overview of existing surface hopping models and LZ formulas.
We then outline the superadiabatic approach and give the resulting formula in Section~\ref{S:superadiabaticFormula},
before describing the associated algorithm in Section~\ref{S:algorithm}.  
In Section~\ref{S:numerics} we systematically investigating its accuracy, and the effects of replacing the true transition probability
by two LZ-like approximations.
Finally, in Section~\ref{S:conclusions}, we summarize our results and discuss some open problems.

\section{The Model}
\label{S:model}

The Schr\"odinger equation governing the quantum dynamics of a molecular system can be 
written as
\[
	\ii \hbar \partial_t \psi(\xn,\xe,t) = H_{\rm mol} \psi(\xn,\xe,t),
\]
where $\xn$ and $\xe$ are the nuclear and electronic positions, respectively, and the Hamiltonian
is given by
\[
	H_{\rm mol} = - \frac{\hbar^2}{2 \mn} \Delta_{\xn}
	-  \frac{\hbar^2}{2 \me} \Delta_{\xe}
	+ V_{\rm n}(\xn) + V_{\rm e}(\xe) + V_{{\rm n}, {\rm e}}(\xn,\xe).
\]
Here the first two terms are the kinetic energies of the nuclei and electrons with masses $\mn$ and $\me$,
respectively.  Note that the masses of the nuclei may all be chosen to be the same by a rescaling of the
nuclear coordinates.  The potentials $V_{\rm n}$ and $V_{\rm e}$ denote the nuclear and electronic 
Coulomb repulsions, respectively, whilst $V_{\rm n, \rm e}$ is the attraction between the nuclei and electrons.

We now change to atomic units ($\hbar = \me = e = 1$) and define $\epsilon = 1/\sqrt{\mn}$ and the 
electronic Hamiltonian for fixed nuclear positions $\xn = \bs{x}$,
\[
	H_{\rm e}(\bs{x}) =
	-  \frac{1}{2} \Delta_{\xe}
	+ V_{\rm n}(\bs{x}) + V_{\rm e} + V_{{\rm n}, {\rm e}}(\bs{x},\cdot).
\]
Suppose that $U^{\pm}(\bs{x})$ are two eigenvalues of the electronic Hamiltonian (i.e.\ two adiabatic potential
energy surfaces) of multiplicity one and well-separated from the rest of the electronic spectrum.  Then, from 
Born-Oppenheimer theory~\cite{Hagedorn80,SpohnTeufel01}, the effective nuclear Schr\"odinger equation is
\begin{equation}
	\ii \epsilon \partial_t \psi(\bs{x},t) = \Big( -\frac{\epsilon^2}{2} \Delta_{\bs{x}} + V(\bs{x}) \Big) \psi(\bs{x},t),
	\label{BOSE}
\end{equation}
where $V$ is a $2 \times 2$ matrix with eigenvalues $U^{\pm}$, i.e.\ a diabatic matrix.
In general, $V$ is symmetric and has the form
\[
	V(\bs{x}) = \begin{pmatrix} V_1(\bs{x}) & V_{12}(\bs{x}) \\  V_{12}(\bs{x}) & V_2 (\bs{x})\end{pmatrix}.
\]

For notational convenience, and to connect back to previous 
work~\cite{BetzGoddardTeufeul09,BetzGoddard09,BetzGoddard11,BetzGoddardManthe16},
we find it useful to define
\[
	Z = (V_1-V_2)/2, \quad
	X = V_{12}, \quad
	d =(V_1+V_2)/2, \quad 
	\rho =\sqrt{X^2+Z^2}
\]
and so 
\[
	V(\bs{x}) = d(\bs{x}) +  \begin{pmatrix} Z(\bs{x}) & X(\bs{x}) \\  X(\bs{x}) & -Z(\bs{x})\end{pmatrix}.
\]
It is easy to see that the adiabatic surfaces are then given by $U^\pm(\bs{x}) = d(\bs{x}) \pm \rho(\bs{x})$
and so $\rho$ is half the energy gap between the two surfaces.

%\subsection{Observables}
%
%The wavepacket itself is not a quantum observable; it cannot be measured experimentally.  However,
%an accurate representation of the wavefunction allows the computation of quantum observables such as
%the expected position, momentum, energy and mass/probability.  For an observable $\mathcal{O}$, the 
%expected value is given by either
%\[
%	\langle \psi | \mathcal{O} | \psi \rangle = \int \dd \bs{x} \psi(\bs{x}) \mathcal{O}(\bs{x}) \psi^*(\bs{x})
%\]
%or
%\[
%	\langle \hat\psi | \mathcal{O} | \hat\psi \rangle = \int \dd \bs{p} \psi(\bs{p}) \mathcal{O}(\bs{p}) \psi^*(\bs{p}).
%\]
%We are principally interested in in the expected position, momentum and mass, which are given by
%\begin{align*}
%	\langle \psi | \bs{x} | \psi \rangle &= \int \dd \bs{x} \, \bs{x} |\psi(\bs{x})|^2 \\
%	\langle \hat \psi | \bs{p} | \hat \psi \rangle &= \int \dd \bs{p} \, \bs{p} |\hat \psi(\bs{p})|^2 \\
%	\langle \psi | 1 | \psi \rangle &= \int \dd \bs{x}  |\psi(\bs{x})|^2 
%	= \langle \hat \psi | 1 | \hat \psi \rangle = \int \dd \bs{p}  |\hat \psi(\bs{p})|^2.
%\end{align*}

\section{Avoided Crossings} 
\label{S:avoidedCrossings}

In the adiabatic representation, an explicit unitary transformation $U_0$ is applied to the system such that
the electronic Hamiltonian is diagonal at each choice of $\bs{x}$.  
Transitions between the adiabatic surfaces are then governed by the kinetic energy term, which introduces
off-diagonal coupling elements, giving (for a one-dimensional (1D) system) a Hamiltonian of the form
\begin{equation}
	H_0 = -\frac{\epsilon^2}{2} \partial_{x}^2 +
	\begin{pmatrix} 
		U^+(x) & - \epsilon \kappa(x) ( \epsilon \partial_x)\\ 
		  \epsilon \kappa(x) ( \epsilon \partial_x) & U^-(x)
	\end{pmatrix}
	+ \mathcal{O}(\epsilon^2).
	\label{adiabaticH}
\end{equation}
Here $\kappa = (X'Z - Z'X)/\rho^2$ is an explicit `kinetic coupling' function and we have grouped the terms such that it is
more obvious that the off-diagonal elements of the above matrix are of order $\epsilon$.  This can be seen
from the fact that wavepackets typically oscillate with frequency $1/\epsilon$ (see Section~\ref{S:avoidedCrossings}), 
so $ \epsilon \partial_x \psi(x)$ is of order one.   Hence we see that, na\"ively, the transitions are of order $\epsilon$
globally in time.   However, as discussed previously, the transitions are exponentially small in $1/\epsilon$ away from the avoided crossings. 

Typically, when the adiabatic potentials are well-separated, the 
coupling elements are small and then two levels may be treated separately via the Born-Oppenheimer 
approximation.  However, if the adiabatic surfaces become close, but do not cross, the coupling terms
typically become large (but do not diverge).  Such nuclear configurations are known as avoided crossings.
As a result of the large coupling elements, a small, but not negligible, part of the nuclear wavepacket is 
transferred between the adiabatic surfaces.

\begin{figure}[h]
	\includegraphics[width = \textwidth]{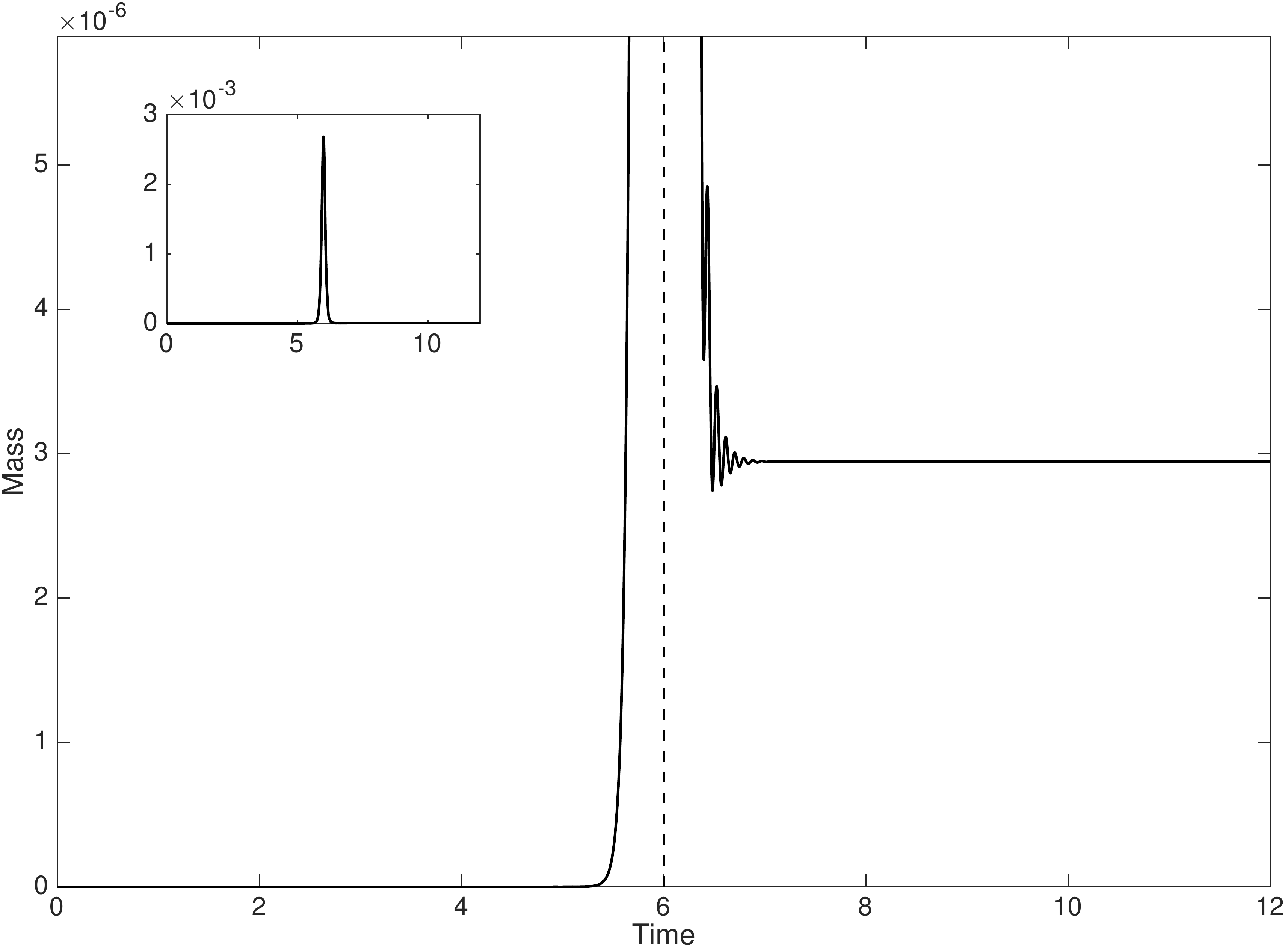}
	\caption{Inset: The mass of the transmitted wavepacket against time as the wavepacket on the original adiabatic
	surface moves through an avoided crossing.  Main figure: Zoom for clarity of the final
	transmitted mass and St\"uckelberg oscillations. The time at which the centre of mass of the original wavepacket 
	reaches the avoided crossing is marked with a dashed vertical line and coincides with the maximum transmitted mass.}
	\label{Fig:MassOscillations}
\end{figure}

Suppose, for clarity of exposition, that the wavepacket initially occupies the upper adiabatic level.
The aim of this work is to determine the transmitted wavepacket (on the lower adiabatic level)
well away from the crossing (in the scattering regime).  Whilst one can, in principle, compute this by a
standard numerical solution of the Schr\"odinger equatiion, there are a number of challenges that 
prevent this from being a realistic option for most systems of interest:
\begin{enumerate}
\item In order to compute the dynamics, one needs an accurate representation of the potential
energy surfaces.  Typically the adiabatic surfaces are calculated using quantum chemistry methods, such as 
Density Functional Theory, but it is computationally
expensive to determine such surfaces, especially when the number of degrees of freedom (dimension of $\bs{x}$) 
is large.  In such cases, it is desirable to design methods that can utilise on-the-fly surfaces, determined only
locally.  Additionally, practical methods for determining surfaces for excited states are still in their infancy,
and one also needs to determine the off-diagonal coupling elements. Finally, we note that diabatic representations
are not unique, and those obtained in two- and multiple-level cases may differ significantly~\cite{BelyaevBarklemDickinsonGadea10}.

\item The wavepackets we wish to compute are highly oscillatory, typically oscillating with frequency of order
$\epsilon^{-1}$ in space.  This can be seen by comparing the kinetic and potential terms in \eqref{BOSE}.  
When using a standard numerical scheme, such as Strang splitting, correctly resolving such oscillations requires 
very fine grids in both position and momentum space.  The curse of dimensionality (for $N$ points in $d$ dimensions,
one requires $N^d$ points) results in such approaches being impractical for all but very small dimensional systems.

\item Away from avoided crossings, the transmitted wavepacket is typically exponentially small in both the gap size $\delta$
and $1/\epsilon$.  This can be seen from the formula \eqref{formula} in Section~\ref{S:formula} or the 
standard LZ transition probabilities \eqref{LZd} and \eqref{LZa}, where $\rho_{x_c} = \delta$.  In contrast, globally in time, the
transitions in the adiabatic representation are of order $\epsilon$, which we have already seen from the 
Hamiltonian~\ref{adiabaticH}.  The necessary cancellations in the transmitted
wavepacket occur through St\"uckelberg oscillations.  See Figure~\ref{Fig:MassOscillations} for an example.
  There are two challenges here.  The first is to correctly resolve
these cancellations, which can require very small time steps.  The second is the more general challenge of computing
an exponentially small quantity; any absolute errors in the numerical scheme must also be exponentially small or they
will overwhelm the desired results.
\end{enumerate}

\section{Existing Approaches and Landau-Zener}
\label{S:existing}

In this section we discuss some existing approaches to calculate the transition probability or the
transmitted wavepacket.

\subsection{Surface Hopping Algorithms}

Here we present a brief overview of surface hopping methods, which are one of the most successful approaches for
simulating nonadiabatic dynamics.  Surface hopping is a mixed quantum-classical approach, where particles are transported
classically on the adiabatic surfaces and hop between them under certain conditions, which simulates the quantum
effects.  A general surface hopping algorithm consists of four steps:
\begin{enumerate}
\item Sampling of the initial condition.
\item Classical evolution via $\dot{x} = p$, $\dot{p} = - \nabla U^\pm(x)$.
\item Surface hopping.
\item Computation of observables.
\end{enumerate}
There are many such schemes, both deterministic and probabilistic and we refer to
~\cite{Tully71,MillerGeorge72,Stine76,Kuntz79,Tully90,HammesSchifferTully94,MullerStock97,FabianoGroenhofThiel08,
FermanianKammererLasser08,LasserSwart08,BelyaevLasserTriglia14,BelyaevDomckeLasserTriglia15} for further details.  

Of particular interest here is the surface hopping step.  Typically this is performed when the gap between the
two adiabatic surfaces is minimal along a classical trajectory.  Whenever such a trajectory reaches a local minimum,
a transition to the other surface is performed with a certain probability, usually derived from a simplified quantum
mechanical model.  The standard approach is to use a LZ formula, which we describe in the next section.
The choice of this hopping probability is the main distinguishing feature of different surface hopping models.

The principal advantage of surface hopping algorithms is their simplicity.  Due to their use of classical dynamics, which 
only require local properties of the potential energy surfaces, the methods can be applied in relatively high dimensions,
using on-the-fly surfaces.  As mentioned previously, such high dimensional systems are beyond the reach of full quantum mechanical methods.

The principal disadvantage is that they lose all phase information, and so cannot treat systems in which 
interference effects are important, or determine observables in which the relative phase of the wavepackets 
on the adiabatic surfaces is required~\cite{BelyaevLasserTriglia14}.  Additionally, they are accurate only when the specified hopping probability is
accurate; we will investigate this in Section~\ref{S:numerics}.

\subsection{The Landau-Zener Formula}

As mentioned previously, in order to compute the transition probability, it is common to use a LZ formula.  Whilst the LZ model
provides a simple formula for the transition probability, it is generally obtained from a one-dimensional, two-level model in the 
diabatic representation.  However, practical applications occur in multiple dimensions and the
potential energy surfaces are usually calculated in the adiabatic representation.  There are a number of formulations
of the LZ probability, including the extension to multiple dimensions in the diabatic 
formalism~\cite{FermanianKammererLasser08}, and versions which only require knowledge of the adiabatic potentials
~\cite{ZhuTeranishiNakamura01,BelyaevLebedev11}.
Here we restrict ourselves to two such formalisms, the first is a diabatic representation.
which requires knowledge of the diabatic matrix elements, whilst the second is an adiabatic representation, which only
requires the gap between the adiabatic potentials.  From now on, we consider only 1D systems; see Section~\ref{S:conclusions}
for some discussion of progress in higher dimensions.

Consider a classical particle with trajectory $\big(x(t), p(t) \big)$ in phase space.  Denote the position where $\rho$ attains 
a minimum by $x_c$, and the momentum of the particle at the corresponding time $t_c$ by $p_c$.  Then the diabatic
LZ transition probability is given by
\begin{equation}
	P_{\rm d} = \exp\Big( - \frac{\pi}{\epsilon} \frac{ \rho(x_c)^2 }{|p_c| \sqrt{ X'(x_c)^2 + Z'(x_c)^2} } \Big).
	\label{LZd}
\end{equation}
The corresponding adiabatic transition probability is given by
\[
	P_{\rm a} = \exp\Big( - \frac{ \pi}{\epsilon} \sqrt{ \frac{\rho(x_c)^3} {\frac{\dd^2}{\dd t^2} \rho(x(t))|_{t=t_c} } } \Big).
\]
If one has knowledge of the diabatic matrix elements, and hence $X$ and $Z$, this can be rewritten as
\begin{equation}
	P_{\rm a} = \exp\Big( - \frac{ \pi}{\epsilon} \frac{\rho(x_c)^2} { |p_c|\sqrt{X'(x_c)^2 + Z'(x_c)^2 + X(x_c) X''(x_c) + Z(x_c) Z''(x_c) } } \Big).
	\label{LZa}
\end{equation}
Note that in the corresponding multidimensional formula~\cite{BelyaevLasserTriglia14}  there is an additional term, which in 1D would be 
$[X(x_c) X'(x_c) + Z(x_c) Z'(x_c)] (U^\pm)'(x_c)$.  However, since an avoided crossing is defined as a minimum of 
$\rho$, and $\rho' = (X X' + Z Z')/\rho$, this term is zero in 1D.

\section{Superadiabatic Representations and the Formula}
\label{S:superadiabaticFormula}

In this section we will briefly review the ideas behind the use of superadiabatic representations to compute the 
transmitted wavepacket and refer the reader to the cited works for more details. 
We will then present a generalisation of a previously-derived formula, which is applicable to 1D avoided crossings
not centred at the origin.

\subsection{Superadiabatic Representations} \label{S:superadiabatic}

Superadiabatic representations were first introduced by Berry~\cite{Berry90,BerryLim93}, 
under the additional approximation that the nuclei
move classically.  More recently this has been extended to the full BOA
~\cite{BetzGoddardTeufeul09,BetzGoddard09,BetzGoddard11,BetzGoddardManthe16}.
As suggested by the name, superadiabatic representations are refinements of the adiabatic representation,
which we described in Section~ref{S:avoidedCrossings}.
In the adiabatic representation, transitions can be very complicated, as demonstrated by the population on the lower
level during a typical transition, see Figure~\ref{Fig:MassOscillations}.  This reliance on large cancellations to leave an exponentially
small wavepacket suggests that the adiabatic representation may not be the ideal frame of reference in which
to study transitions at avoided crossings.

Superadiabatic representations improve on the adiabatic one by simplifying the dynamics near an avoided
crossing, at the expense of introducing computational complexities.  
The superadiabatic representations can be enumerated, and, initially, moving to successively higher
superadiabatic representations reduces the spurious oscillations in the dynamics until the transmitted population
builds up monotonically as the wavepacket travels through the avoided crossing.  This is known as the 
optimal superadiabatic representation.  However, moving to even higher representations results in the spurious
oscillations returning. Previous results give a reliable method to determine the optimal superadiabatic 
representation~\cite{BetzGoddardTeufeul09,BetzGoddard11}.  
However, computing the
unitary operators for this representation is highly challenging, and performing the numerical computations in 
such a representation is similarly difficult.

The main benefit of superadiabatic representations for our purposes is that they allow the derivation of an explicit formula
for the transmitted wavepacket in the optimal superadiabatic representation, without requiring the associated unitary
matrix.  By general theory~\cite{Teufel03}, all of the superadiabatic representations agree with the adiabatic one away from
any avoided crossing.  This leads to a simple algorithm to compute the transition through an avoided 
crossing in the adiabatic representation, as described in Section~\ref{S:algorithm}.

\subsection{The Formula} \label{S:formula}

Following Ref.~\cite{BerryLim93}, it is useful to introduce a nonlinear rescaling in
which the adiabatic coupling elements obtain a universal form, known as the natural scale,
\[
	\tau(x) = 2 \int_{x_c}^x \rho(s) \dd s,
\]
where $x_c$ is the position of the avoided crossing.  We now extend $\rho$ and $\tau$ into the complex plane and,
by the theory of Stokes lines~\cite{JoyeMiletiPfister91}, 
the analytic continuation of $\rho$ has a pair of complex conjugate zeros, close to $x_c$, 
at $x_{cz}$ and $x_{cz}^*$.  We define
\[
	\tau_{x_c} = \tau(x_{cz}) = \tau_r + \ii \tau_c.
\]

Let $\phi^\pm(x,t_c)$ be the incoming wavepacket on the corresponding adiabatic surface $U^\pm$ at time $t_c$ 
when the centre of mass coincides with an avoided crossing at $x_c$.  Then, for $t>t_c$, the transmitted wavepacket
on the other adiabatic surface $U^\mp$ can be approximated by
\[
	\psi(x,t) = \e{-(\ii/\epsilon)(t-t_c)H^\mp} \psi^\mp(x)
\]
where $H^\mp$ are the BOA Hamiltonians for the two levels and 
$\psi^\mp(x)$ is a wavepacket instantaneously created at time $t_c$, which is more easily expressed in 
Fourier space via
\begin{align}
	\hat{\psi}^\mp(p) &= \Theta(p^2 \mp 4 \delta) \frac{p + \eta^\mp}{2 |\eta^\mp|} 
					\exp \Big( -\frac{\tau_c}{2 \delta \epsilon} |p - \eta^\mp| \Big)
					\exp \Big( -\ii \frac{\tau_r}{2 \delta \epsilon} (p - \eta^\mp) \Big) \notag \\
					& \qquad \times \exp \Big( -\ii \frac{x_c}{\epsilon} (p - \eta^\mp) \Big)
					\hat{\phi}^\pm(\eta^\mp). 
	\label{formula}
\end{align}
Here
\[
	\delta = \rho(x_c), \quad \eta^\mp = \sign(p) \sqrt{p^2 \mp 4 \delta},
\]
$\Theta$ is the Heaviside function and the Fourier transform needs to be performed under the
correct scaling:
\[
	\hat{\psi}(p) = \frac{1}{2\pi \epsilon} \int \e{- (\ii/\epsilon) p x} \psi(x) \dd x.
\]

We note that the principle difference from previous presentations of the formula is the final exponential
factor involving $x_c$, the position of the avoided crossing.  In previous work, this position has been taken to be zero,
in which case the factor is simply 1.  The new term arises from the approximation $x(\tau) = \tau/(2\delta) + x_c + \mathcal{O}(\tau^3)$ 
(which is a simple generalisation of the calculation for $x_c=0$~\cite[p.\ 2258]{BetzGoddard11}).

\subsection{Analysis of the Formula}\label{S:formulaAnalysis}

We now present a brief analysis of the formula in~\ref{formula}, which allows us to connect to surface hopping approaches,
as well as to LZ formulas.

Firstly we note that the formula involves the same momentum adjustment that is phenomenologically introduced in
surface hopping algorithms.  
We note that $\eta\mp$ is precisely the classical incoming momentum required to give outgoing momentum $p$
when moving down/up, respectively, a potential energy gap of $2\delta$ and requiring (classical) energy conservation.
Relatedly, when passing from the upper to the lower level, the Heaviside function ensures that the
transmitted wavepacket has (absolute) momentum at least $2 \delta$, whereas when passing from the lower to upper 
level it is trivially 1, indicating no restriction on the transmitted momentum.  The analogous restriction that a 
classical particle can only be transmitted to the upper level if it has sufficient kinetic energy is accounted for by the
$\hat{\psi}^\pm(\eta^\mp)$ term.

We now discuss how, in appropriate limits, the formula essentially reduces to a LZ transition for each point in momentum
space.  We make a number of independent approximations:
\begin{enumerate}
	\item $x_c = 0$.\\
	For a single avoided crossing we may do this without loss of generality by shifting the space variable.

	\item $\tau_r = 0$.\\  
	This is the case, for example, when the potential is symmetric around the avoided crossing.

	\item $\delta$ is small.\\  
	This produces two simplifications to the formula using that $\eta^\mp \approx p \mp 2\delta/p$:
	\begin{itemize}
		\item The prefactor simplifies to $\frac{\eta^\mp + p}{2|\eta^\mp|} \approx p/|p| = \sign(p)$;
		\item The factor in the exponential simplifies to $|p - \eta^\mp| \approx 2\delta / |p|$.
	\end{itemize}
	Note that the small parameter in these expansions is actually $\delta/p_0$, and so we expect these approximations
	to be more accurate for either small $\delta$ or large incoming momentum.

	\item Second order expansion of $\rho$.\\
	It is well known~\cite{BerryLim93} that a na\"ive second order expansion of $\rho$ is incorrect as the analytic
	continuation of $\rho$ must vanish like a square root at its complex zeros.  We therefore approximate $\rho$
	via
	\[
		\rho(x) \approx \sqrt{\delta^2 + g(x-x_c)}
	\]
	with $g$ a smooth function such that $g(0) = g'(0) =0$.  Performing a second order expansion of $g$ then gives
	\[
		\rho(x) \approx \sqrt{ \delta^2 + \tfrac{1}{2} g''(0) (x-x_c)^2}.
	\]
	In this case both $x_{cz}$ and $\tau_c$ can be computed analytically to give
	\[
		\tau_c \approx \ii \frac{\pi \delta^2}{2 \alpha}
	\]
	where $\alpha^2 = \tfrac{1}{2} g''(0)$.  To connect purely to $\rho$, we note that $\frac{1}{2} g''(0) = \delta \rho''(x_c)$
	and hence
	\[
		\tau_c \approx \ii \frac{\pi \delta^{3/2}}{2 \sqrt{\rho''(x_c)}}.
	\]
	Finally, in order to connect to the LZ formulas, an explicit computation using that 
	$\rho'(x_c) = (X(x_c) X'(x_c) + Z(x_c) Z'(x_c))/\rho(x_c) = 0$ and $\rho(x_c) = \delta$
	gives 
%	\[
%		\rho''(x) = \frac{X'(x)^2 + Z'(x)^2 + X(x) X''(x) + Z(x) Z''(x)}{\rho(x)}
%				- \frac{ \big( X(x) X'(x) + Z(x) Z'(x) \big)^2}{\rho(x)^3}
%	\]
	\[
		\rho''(x_c) = \frac{X'(x_c)^2 + Z'(x_c)^2 + X(x_c) X''(x_c) + Z(x_c) Z''(x_c)}{\delta}
	\]
	and so
	\[
		\tau_c \approx \ii \frac{\pi \delta^{2}}{2 \sqrt{X'(x_c)^2 + Z'(x_c)^2 + X(x_c) X''(x_c) + Z(x_c) Z''(x_c)}}.
	\]
	
\end{enumerate}

Suppose now that we make all four approximations.  Then the formula in~\eqref{formula} becomes
\begin{equation}
	\hat{\psi}^\mp(p) = \sign(p) \Theta(p^2 \mp 4 \delta) 
					\exp \Big( - \frac{\pi \delta^{2}}{2 \epsilon |p| \sqrt{X'(x_c)^2 + Z'(x_c)^2 + X(x_c) X''(x_c) + Z(x_c) Z''(x_c)}}  \Big)
					\hat{\psi}^\pm(\eta^\mp).
	\label{LZformula}
\end{equation}
It is now clear that the exponential factor corresponds precisely to the adiabatic LZ transition probability
in \eqref{LZa}, with the additional factor of $1/2$ accounting for the fact that we are determining the size of the transmitted
wavepacket rather than the transition probability, which is proportional to the square of the wavepacket.
The Heaviside function is also included indirectly in surface hopping models, which explicitly exclude 
classically-forbidden transitions, see e.g.~\cite{BelyaevLasserTriglia14}.

Note that if we are interested solely in the transition probability then the first two approximations are 
irrelevant as they only affect the phase.  However, when dealing with multiple transitions these terms are
crucial in understanding interference effects.  In Section~\ref{S:numerics} we will investigate the effects of these
approximations in some example systems.

After approximations (1)--(4) have been made, the resulting formula~\eqref{LZformula} can be thought of as a surface
hopping algorithm that retains phase information.  This can be seen by noting that the formula decouples in momentum space.
Thus, if we replace the classical transport of individual particles, the ensemble of which represents the initial wavepacket,
with quantum evolution of the initial wavepacket, and then replace particle hopping with hopping of momentum components
of the wavepacket, then we have a clear analogue of the surface hopping methods.  One promising avenue of further work
is to investigate the use of formula~\ref{formula} for the transmission probability (instead of the LZ one) in 
traditional surface hopping algorithms.
Alternatively, we can recover the surface hopping methodology (but retaining phase information) 
by dividing the wavepacket into small pieces (the surface hopping particles), evolving them classically
on the initial level (e.g.\ using Hagedorn's wavepacket approach~\cite{Hagedorn81,Hagedorn94,Lubich08})
until they reach an avoided crossing, and then applying the formula either with the full transition
probability, or the LZ approximation, and reconstructing the wavepacket on the other level.

\section{The Algorithm} 
\label{S:algorithm}

The general algorithm described below is similar to that presented in previous work, but here it is 
extended to multiple transitions and to different levels of approximation, which ultimately lead
to an analogue of the LZ formula, but applied to wavepackets, rather than simply as
a transition probability.  The transmitted wavepacket is computed via the following algorithm.
For clarity, we present the algorithm for two BOA surfaces, but its extension to multiple surfaces
is straightforward due to the linearity of the Schr\"odinger equation.

\begin{enumerate}
\item \textbf{Initial Condition:}  The initial wavepacket should be specified on either the upper or lower
adiabatic level, well away from any of the avoided crossings.  Note that, in such regions, the adiabatic, superadiabatic,
and diabatic levels are very close, so one may instead specify the wavepacket on a single diabatic level.
If the initial wavepacket is given close to an avoided crossing, for example as the result of a laser excitation,
then it must be evolved away (into the scattering regime) on the corresponding adiabatic level under the BO approximation
to obtain an appropriate initial wavepacket.

\item \textbf{One-Level Dynamics:}  The initial wavepacket is now evolved under the BOA
until the final, specified time, or until another termination condition is satisfied
(such as the wavepacket reaching a minimum distance from an avoided crossing).  This can be done using any
one-level scheme that provides sufficient accuracy, such as Strang splitting, Hagedorn wavepackets~\cite{Hagedorn81,Hagedorn94,Lubich08}, or MCTDH~\cite{MCTDHBook}.  
The wavepacket on the other BOA level is evolved simultaneously; the level is initially unoccupied.

\item \textbf{Detection of Avoided Crossings:}  Here an avoided crossing is defined as a (local) minimum of 
the gap $\rho$.
Whenever the centre of mass of the wavepacket reaches such a minimum, apply the formula as described in the next step.
Such local minima may be determined \emph{a priori}, for example when the potentials are given analytically,
or on-the-fly by monitoring $\rho$.

\item \textbf{Application of the Formula:}  Apply the formula~\eqref{formula} to the wavepacket at the avoided crossing and add the 
resulting wavepacket to the lower level.  Note that the formula implicitly requires the potentials to be extended into
the complex plane in order to compute $\tau$.  However, as described in the following Section, this requirement 
may be bypassed by using an analogue of the Landau-Zener formula, at a cost to accuracy which is investigated
for some examples in Section~\ref{S:numerics}.

\item \textbf{Computation of Observables:} At any time step the wavepackets on the two levels may be used to 
compute observables, such as mean position, momentum and the level populations, including those which require
phase information such as inter-level observables.  Note, however (as discussed
in Section~\ref{S:superadiabatic}), that these will only agree with the corresponding quantities computed for the adiabatic populations
well away from any avoided crossings.  An extreme example of this is that, before the wavepacket on the initial level
reaches the avoided crossing, the other level us completely unoccupied; see Figure~\ref{Fig:QuadMass}.

\end{enumerate}

In the following section we will investigate the accuracy of this algorithm.  One restriction for its application to 
multiple crossings is that the transmitted wavepacket must be small, or, more precisely, the wavepacket remaining 
on the original surface must not change significantly when compared to its evolution under the BOA.
This is due to the perturbative nature of the derivation, which assumes that the original wavepacket is unchanged
during a transition.

\section{Numerics} \label{S:numerics}

Note that the MATLAB code used to produce the results in this section is available from \url{https://bitbucket.org/bdgoddard/qmd1dpublic/}.

\subsection{Jahn-Teller}

We consider first a simple example in order to demonstrate the effects of the approximations in Section~\ref{S:formulaAnalysis}.
We choose
\[
	V(x) = \begin{pmatrix} x & \delta \\ \delta & -x \end{pmatrix}
\]
where we have $X = \delta$, $Z = x$, $\rho = \sqrt{x^2 + \delta^2}$.  There is a single avoided crossing at
$x_c = 0$, with gap $2\delta$.  It is clear that $x_{cz} = \ii \delta$ and a straightforward calculation shows that
$\tau_{x_c} = \ii \delta^2 \pi/2$.  Note, therefore, that assumptions (1), (2) and (4) of Section~\ref{S:formulaAnalysis} hold exactly.
Furthermore, since $X(x_c)X''(x_c) + Z(x_c)Z''(x_c) =0$, the diabatic and adiabatic LZ transition probabilities
given in \eqref{LZd} and \eqref{LZa} are identical in this case.  This simple model allows us to investigate the
effects of approximation (3), i.e.\ the difference between the full formula \eqref{formula} and the LZ approximation
for a range of values of $\delta$.  From the arguments in Section~\ref{S:formulaAnalysis}, we expect the two results to agree
to high accuracy when $\delta/p_0$ is small, and hence the transition is large, but we expect the full formula result to be
more accurate in the regime of interest (relatively large $\delta$ and small transitions).  

\begin{figure}
\includegraphics[width=0.49\textwidth]{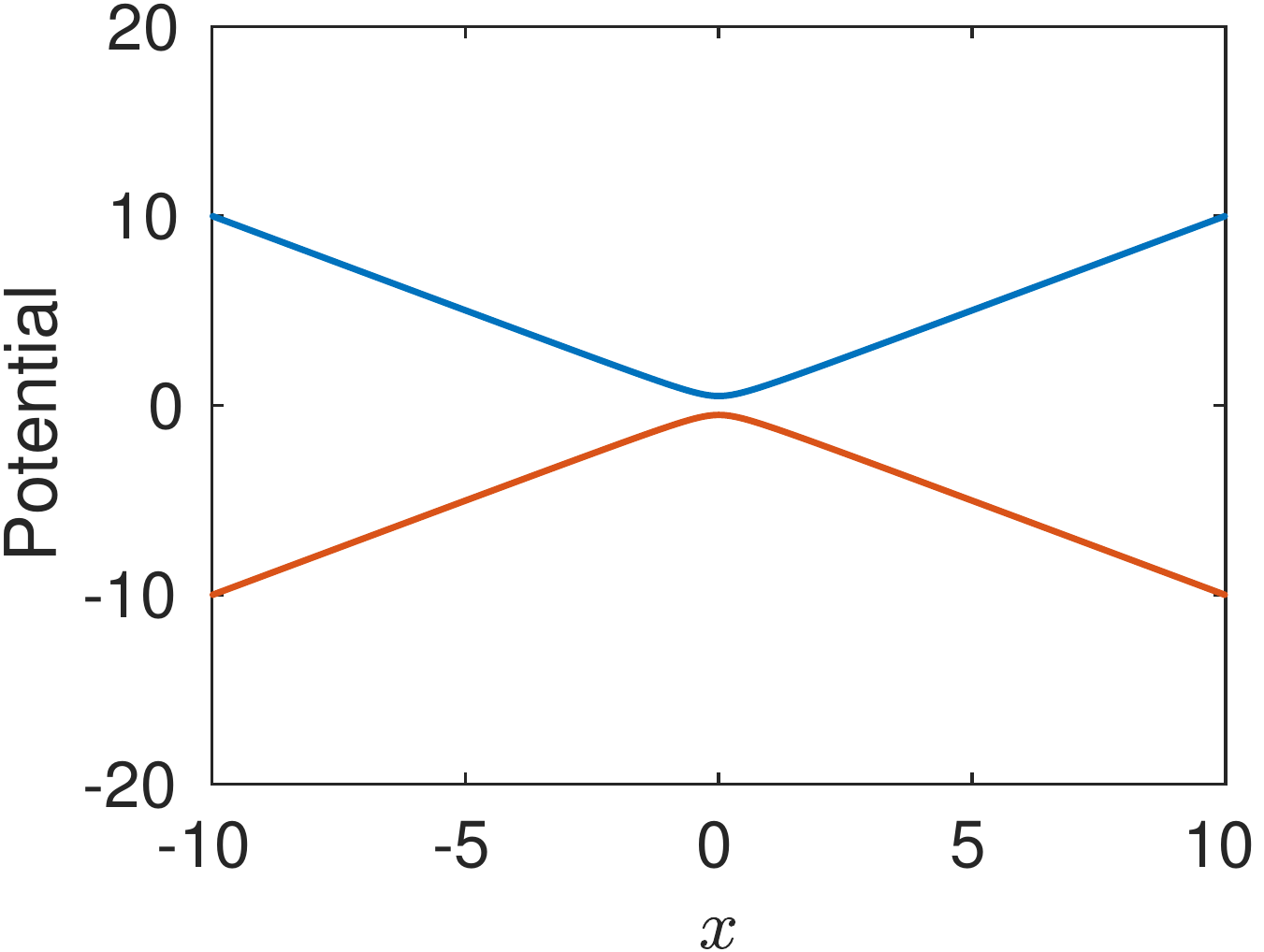}
\includegraphics[width=0.49\textwidth]{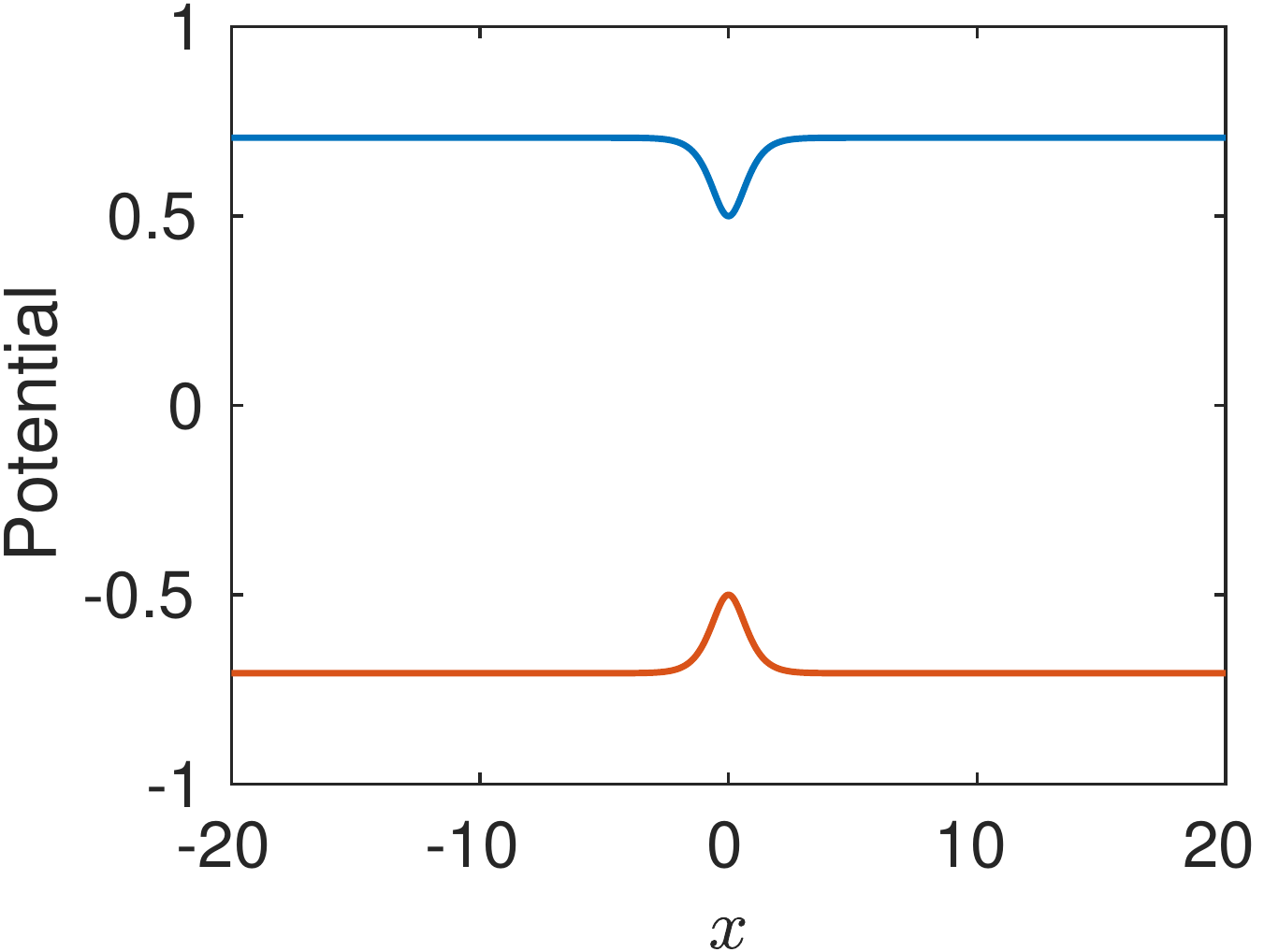}\\
\includegraphics[width=0.49\textwidth]{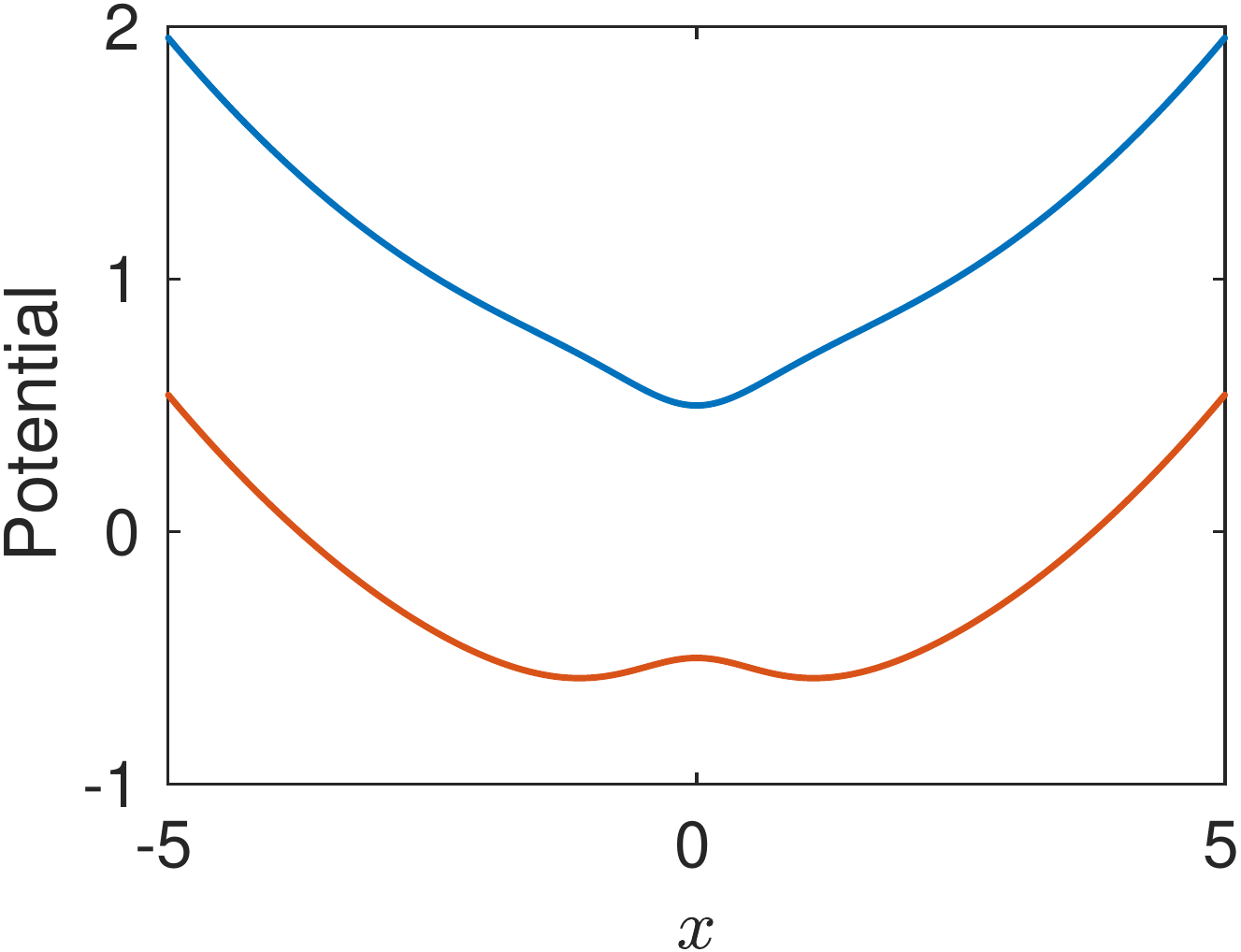}
\includegraphics[width=0.49\textwidth]{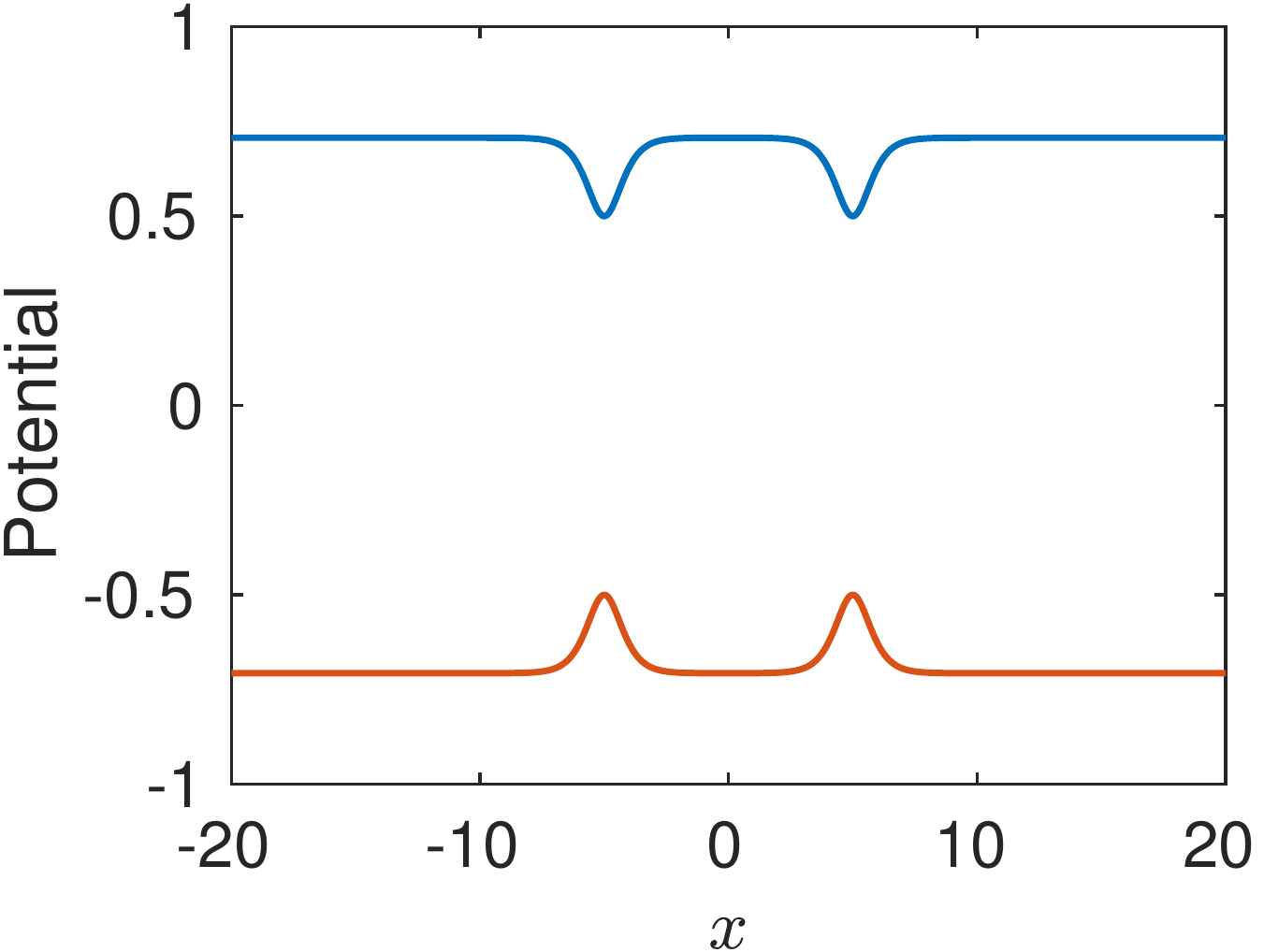}
\caption{The adiabatic potentials for [clockwise from top left] the Jahn-Teller, Simple, Dual
and Quadratic Potentials}.
\label{Fig:Potentials}
\end{figure}

We choose to specify the wavepacket at the avoided crossing, and determine the initial condition by evolving it
backwards in time away from the crossing on a single adiabatic surface.  This ensures that the wavepacket is semiclassical
(i.e.\ of width order $\sqrt{\epsilon}$) when it reaches the avoided crossing.  As noted above, due to the linearity of
the Schr\"odinger equation, if this were not the case then we could apply a slicing procedure to obtain similarly
accurate results.  In particular, we choose
\begin{equation}
	\hat\psi(p) = \frac{1}{(\pi \epsilon)^{1/4}} \exp \Big( -\frac{\ii}{\epsilon} p_0 x_0 - \frac{1}{2 \epsilon} (p-p_0)^2 
											- \frac{\ii}{\epsilon} x_0 (p-p_0) \Big)
	\label{GaussianP}
\end{equation}
where, along with $\delta$, the free parameters are $\epsilon$ and $p_0$.  For this example we fix $\epsilon = 1/50$,
which is similar to the value chosen in surface hopping 
works e.g.~\cite{FermanianKammererLasser08,BelyaevLasserTriglia14,FermanianKammererLasser17} (and approximately
correct for real-world systems e.g.~\cite{BetzGoddardManthe16})
and $p_0 = 8$.  We could, in principle, vary these parameters, and 
we will do so in later examples.  We note that due to the nature of the potential, in order to start sufficiently
far away from the avoided crossing (such that the adiabatic and superadiabatic representations agree)
the initial potential energy must be reasonably large, leading to a correspondingly large minimum value
of $p_0$ at the avoiding crossing.  See Figure~\ref{Fig:Potentials} for the adiabatic potentials.

Here we evolve backwards to a start time of $-40/p_0$ with timestep $1/(1000 p_0)$ and then forwards through
the avoided crossing for time $80/p_0$ with the same timestep.  We perform the numerics with a spatial grid with
$2^{15}$ points and endpoints $\pm 60$. 
We observe that halving the time step and doubling the number of grid points does not significantly affect the results.

\begin{figure}
\includegraphics[width=0.49\textwidth]{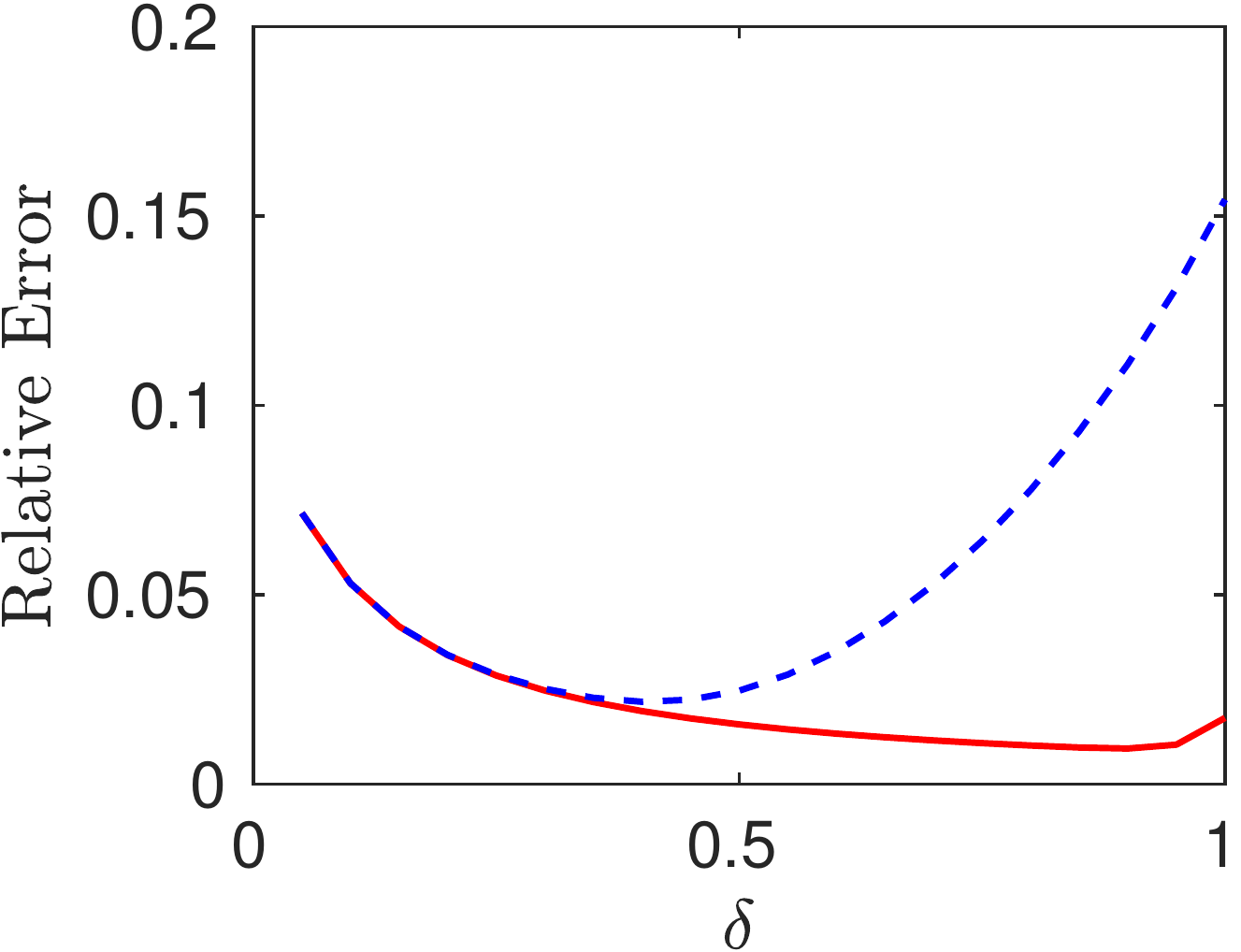}
\includegraphics[width=0.49\textwidth]{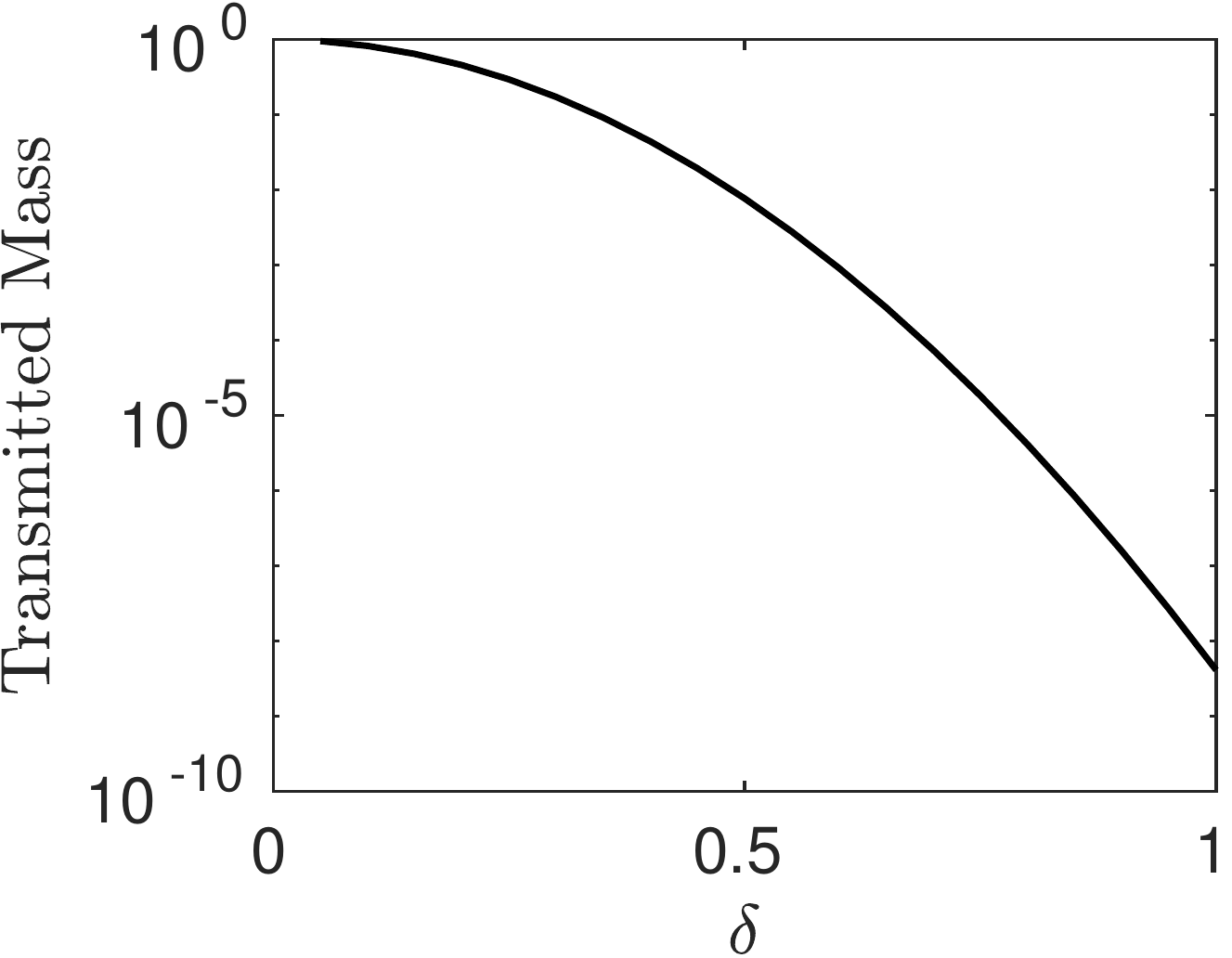}
\caption{Left: The relative error between the `exact' numerical solution and the application of the algorithm using 
formula~\eqref{formula} [red, solid] and~\eqref{LZformula} [blue, dashed].  Right: The `exact' transmitted mass, which
is in excellent agreement with that computed using~\eqref{formula} for all values of $\delta$.}
\label{Fig:JT}
\end{figure}

In Figure~\ref{Fig:JT} we show the relative error in the transmitted wavepacket and the transmitted mass.  This 
clearly demonstrates that, for small $\delta$ (and large transmitted mass), both our formula~\eqref{formula} and
the LZ-like version~\eqref{LZformula} give very good results.  However, as $\delta$ increases, the
simplified version becomes increasingly inaccurate.

\subsection{Simple Avoided Crossing} \label{S:simple}

We now consider a simple example, which will both allow us to systematically investigate the accuracy of our method
for different parameter regimes, and provide a benchmark for the accuracy of a single transition; this is, at least
heuristically, a lower bound for the accuracy for multiple transitions. We choose
\[
	V(x) = \begin{pmatrix} \tfrac{1}{2} \tanh(x) & \delta \\ \delta & - \tfrac{1}{2}\tanh(x) \end{pmatrix}
\]
where we have $X = \delta$, $Z = \tfrac{1}{2}\tanh(x)$, $\rho = \sqrt{\tfrac{1}{4}\tanh(x)^2 + \delta^2}$. 
See Figure~\ref{Fig:Potentials} for the adiabatic potentials with $\delta = 1/2$.

\begin{figure}
\includegraphics[width=\textwidth]{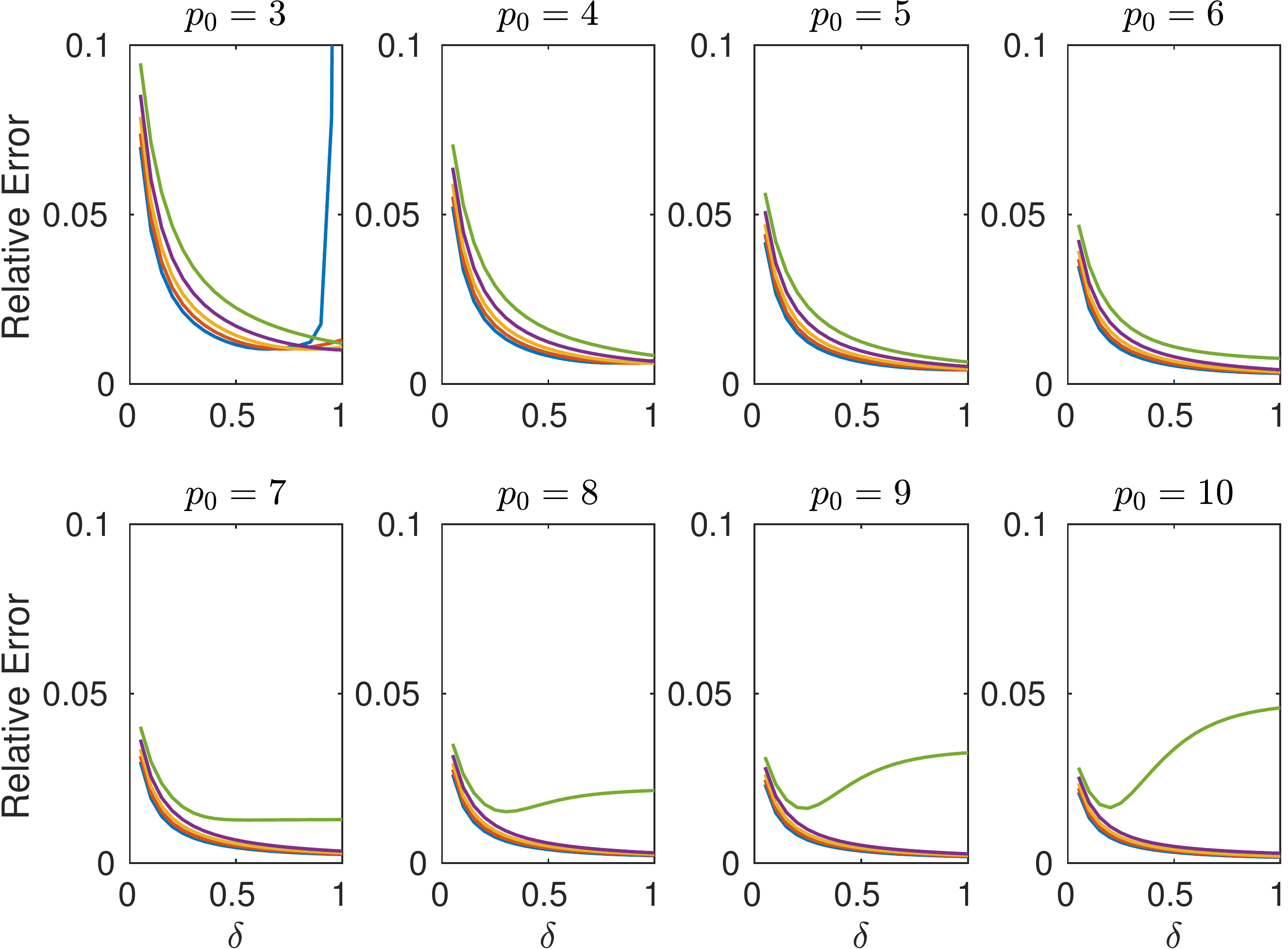}
\caption{The relative error between the `exact' numerical solution and the application of the algorithm using 
formula \eqref{formula}.  Each subplot shows the result for a different value of $p_0$ for a range of $\delta$ values.
Different colour curves $\{$green, purple, yellow, red, blue$\}$ correspond to 
$\epsilon = \{ 1/10, 1/20, 1/30, 1/40, 1/50\}$, respectively.  Note that, apart from the largest value $\epsilon = 1/10$,
the errors are very similar.}
\label{Fig:SimpleFormula}
\end{figure}

As in the previous example, in order to control the (mean) momentum of the wavepacket when it reaches the crossing, 
we specify the wavepacket 
in momentum space at the avoided crossing and then evolve it backwards in time on a single adiabatic surface 
to obtain an initial wavepacket for the computations.  In particular, we take a Gaussian wavepacket as given in 
\eqref{GaussianP} for a range of values of $\epsilon$ and $p_0$.  We compute the results for a single transition of the
avoided crossing, both using the full formula \eqref{formula} and the LZ-like one \eqref{LZformula}.  
As can be seen from Figure~\ref{Fig:SimpleFormula}, the relative
error when using~\ref{formula} is typically of the order of a few percent, with increasing accuracy as $\delta$ and/or $p_0$ increase.  The
deviation of the green curve, which corresponds to $\epsilon = 1/10$ is a result of the asymptotic nature of the 
formula.  The odd behaviour of the blue curve for $p_0=3$, $\epsilon = 1/50$ and $\delta \approx 1$ 
seems to be a result of parts of the wavepacket
becoming `trapped' near the avoided crossing, which violates the assumption of a single transition.

\begin{figure}
\includegraphics[width=\textwidth]{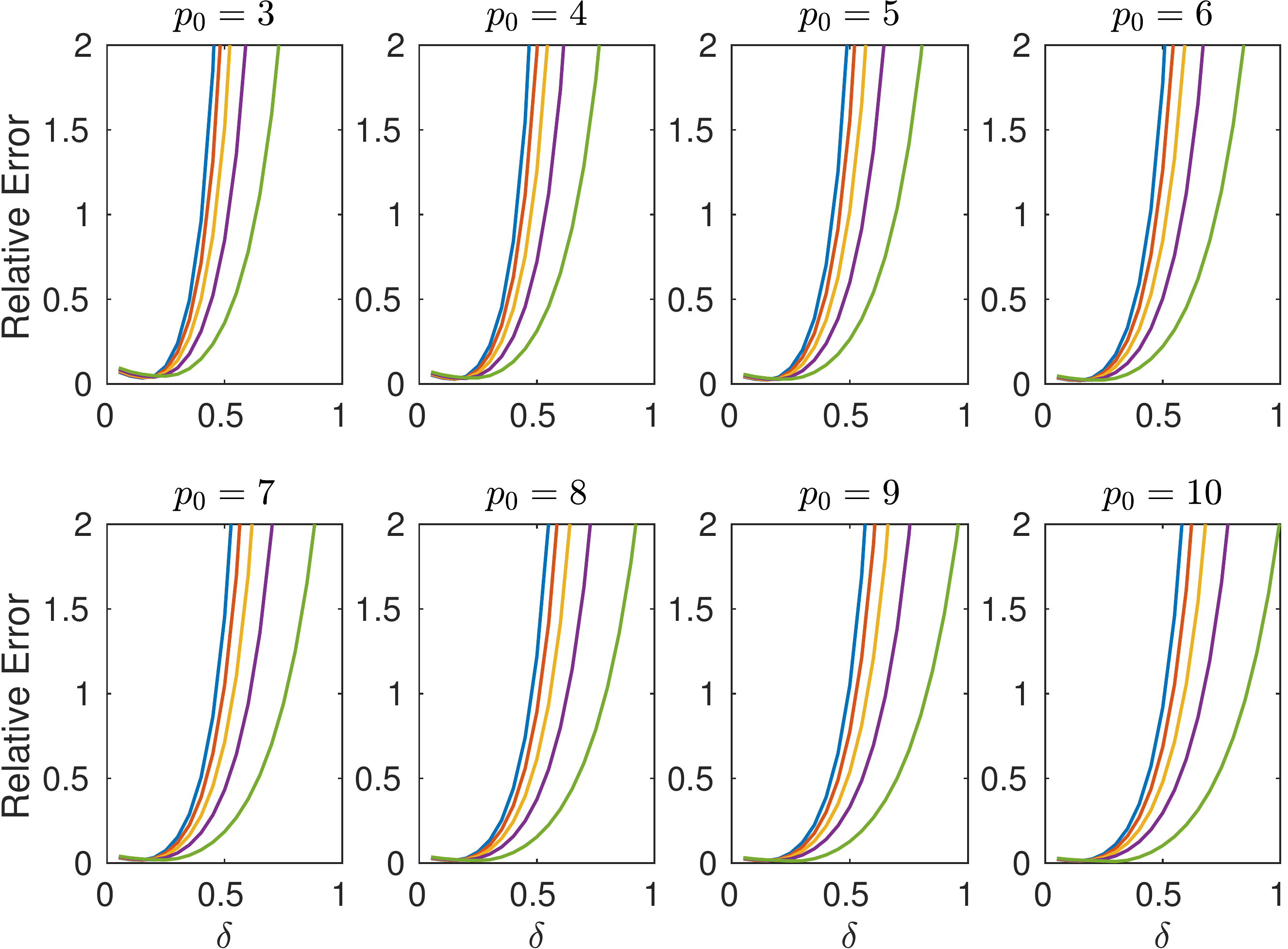}
\caption{As Figure~\ref{Fig:SimpleFormula} but using formula \eqref{LZformula}.   Note that the results for all but
the smallest values of $\delta$ are significantly worse than those in Figure~\ref{Fig:SimpleFormula}.}
\label{Fig:SimpleLZ}
\end{figure}

\begin{figure}
\includegraphics[width=\textwidth]{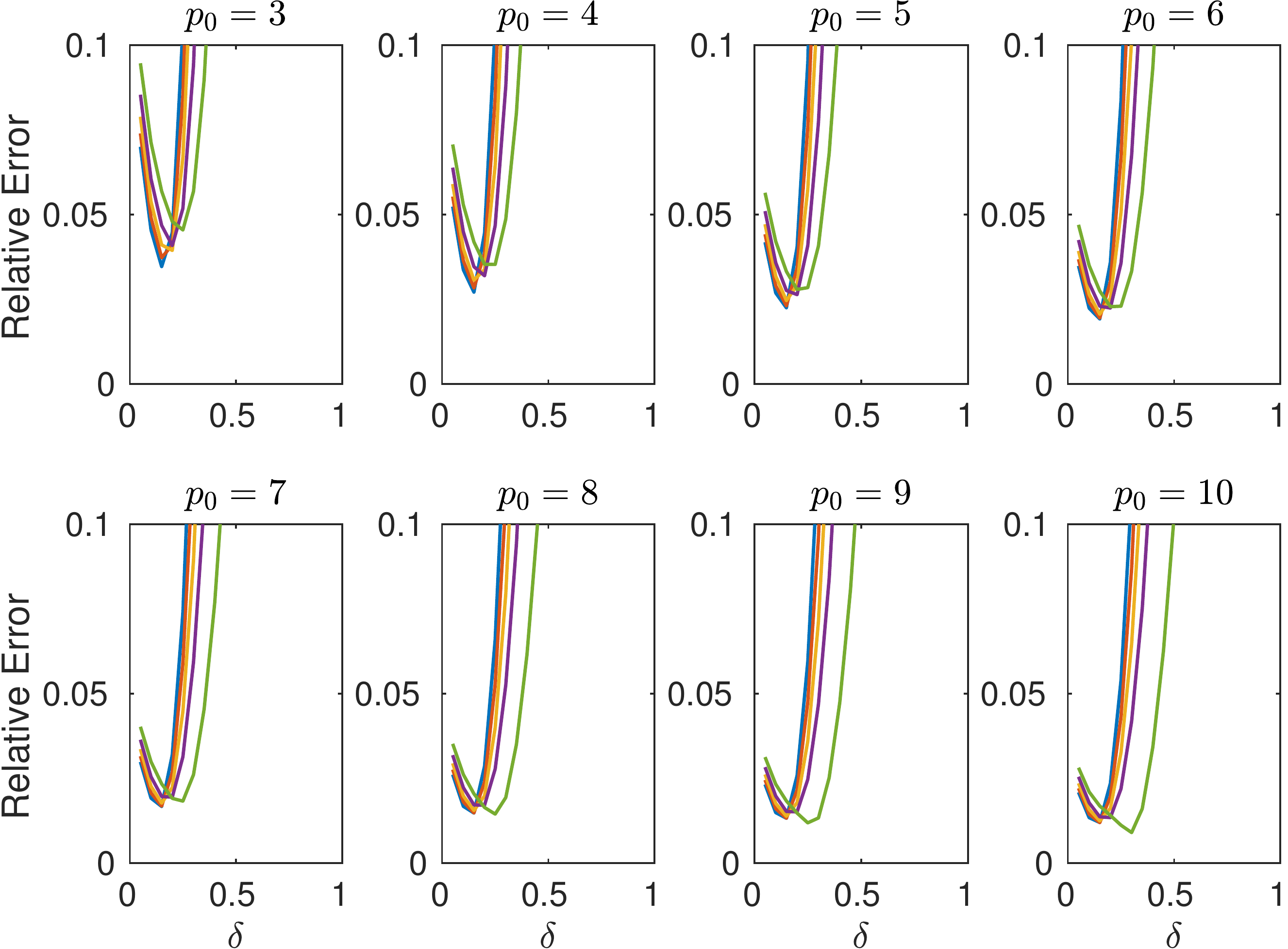}
\caption{Zoom of Figure~\ref{Fig:SimpleLZ}.}
\label{Fig:SimpleLZZoom}
\end{figure}

Figures~\ref{Fig:SimpleLZ} and \ref{Fig:SimpleLZZoom} demonstrate the effects of using the algorithm with the approximate
formula~\eqref{LZformula}.  As can be seen from Figure~\ref{Fig:SimpleLZ}, for moderate values of $\delta$, the results
become very poor.  However, as expected, Figure~\ref{Fig:SimpleLZZoom} shows that, for small $\delta$, 
the results are very similar to those using the full formula~\eqref{formula}.

For completeness, we give the numerical details:  The spatial grid uses $2^{14}$ points with limits $\mp 40$.
We use a time step of $1/(100 p_0)$ and obtain the initial wavepacket by evolving the wavepacket backwards 
from the crossing for time $20/p_0$.  The system is then evolved forwards for time $40/p_0$.  Again, we note that
halving the time step and doubling the number of grid points does not significantly affect the results.

\begin{figure}
\includegraphics[width=\textwidth]{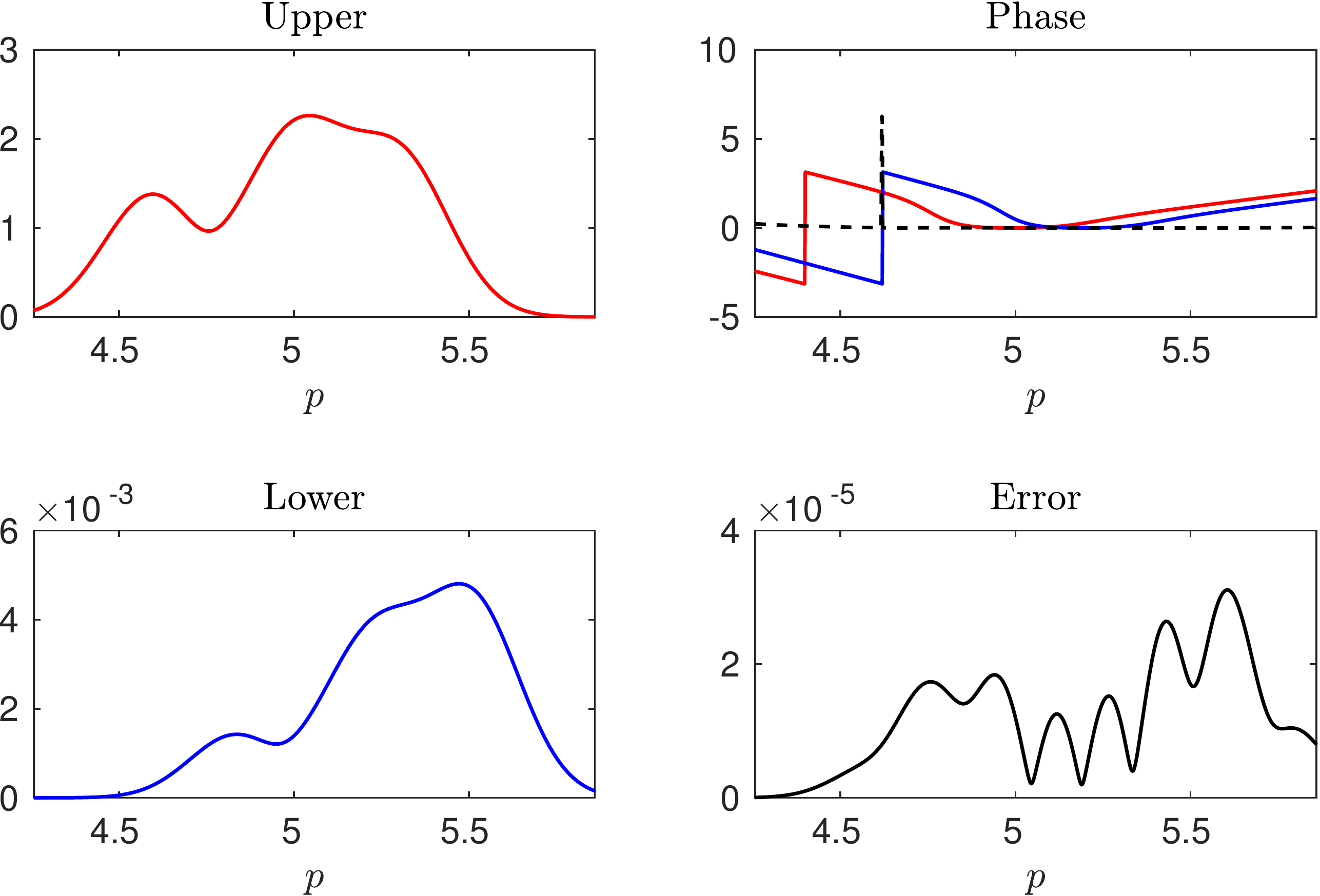}
\caption{The wavepacket in momentum space.  `Upper' denotes the wavepacket on the upper level at the 
avoided crossing, as given by~\eqref{psiHatNG}.  `Lower' denotes the transmitted wavepacket, computed using
the algorithm, evolved backwards on the lower level to the avoided crossing.  `Phase' shows the phase of the
upper (red), lower (blue) and error, for the lower, transmitted phase (black, dashed).  
`Relative Error' displays the relative error between the 
transmitted wavepackets given by the `exact' solution and the result of the algorithm.}
\label{Fig:NG}
\end{figure}

As a further test of the accuracy  of the algorithm we perform the same calculation as for the Gaussian wavepackets
in the previous example, but with a wavepacket on the upper level at the crossing given by
\begin{equation}
	\hat\psi(p) = \sum_{j=1}^3 w_i \hat\psi(x_{0,i},p_{0,i},p),
	\label{psiHatNG}
\end{equation}
where $\hat\psi$ is a Gaussian as given by~\eqref{GaussianP}.  We choose $\epsilon = 1/50$, 
$w = [0.7,1,0.9]$, $p_0 = [4.6,5,5.3]$ and $x_0 = [0.1,0,-0.05]$.  However, we note that the results are robust under
these choices for a wide range of values.  We show the resulting transmitted wavepacket in Figure~\ref{Fig:NG}
which for convenience of displaying the phase, we have evolved backwards to the avoided crossing on the lower level.
Note that the relative error in this case is 0.0057.  In particular, Figure~\ref{Fig:NG} demonstrates that higher-momentum
wavepackets are more likely to make the transition.

We note here that the results for wavepackets starting on the lower level are very similar, and we will investigate
such a situation in the following Section.  

\subsubsection{Full Crossings}

Here we consider if the algorithm is applicable to full crossings (with $\delta=0$).  In such a case, the approximations
made in Section~\ref{S:formulaAnalysis} lead to the conclusion that the transmitted wavepacket is approximately
equal to the incoming wavepacket.  Applying this in the case $p_0=5$, $\epsilon=1/50$ and $\delta=0$ gives a relative
of 0.0856 for both the formula~\eqref{formula} and LZ approximation~\eqref{LZformula}.  This is comparable to the 
relative error for small, but non-zero $\delta$ (see Figure~\ref{Fig:SimpleFormula}).  This indicates that the methodology
can also be used for full crossings.  This is important in higher dimensions, where part of the wavepacket may travel across 
a full crossing (conical intersection), whilst other parts experience an effective avoided crossing, in which case we need
only one algorithm to accurately treat the whole wavepacket.

\subsubsection{Diabatic vs.\ Adiabatic LZ}

In the previous examples, we have $X(x_c)X''(x_c) + Z(x_c)Z''(x_c) = 0$ and hence the 
diabatic and adiabatic LZ transition probabilities, \eqref{LZd} and \eqref{LZa}, respectively, are identical.  
However, here we briefly consider an example in which this is not the case.  We note that such a situation
was also investigated in Ref.~\cite{BelyaevLasserTriglia14} for two-dimensional crossings and the results when using
the two formalisms were found to be very similar.  We will now show that this is not always the case.  We take
a sightly perturbed version of the simple potential matrix above
\[
	V(x) = \begin{pmatrix} \tfrac{1}{2} \tanh(x) & \delta + \tfrac{1}{10} \tanh^2(x)\\ \delta + \tfrac{1}{10} \tanh^2(x) & - \tfrac{1}{2}\tanh(x) \end{pmatrix},
\]
i.e\ $X = \delta + \tfrac{1}{10} \tanh^2(x)$ and $Z = \tfrac{1}{2}\tanh(x)$.  We choose $\delta = 0.2$,
$\epsilon = 1/50$ and $p_0 = 5$ where these parameters, 
are chosen such that we are in a regime where we expect both the formula~\eqref{formula}
and the (adiabatic) LZ approximation to be reasonably accurate, whilst simultaneously the results are not dominated
by $\delta$ being very small.   We use the same numerical scheme as for the simple crossing above and, 
find that the relative errors when using the formula~\eqref{formula}
and the adiabatic LZ approximation~\eqref{LZa} (or~\eqref{LZformula}) are very similar at 0.0219 and 0.0217, respectively.
In contrast, when using the diabatic approximation~\eqref{LZd}, the results are much worse, with a relative error of
0.1081.  This, along with previous results, demonstrates a clear motivation to use the transition formula
in surface hopping algorithms.

\subsection{Multiple Transitions of a Single Crossing}

We now demonstrate the algorithm when the wavepacket makes multiple transitions of a single avoided crossing.
Here we add a quadratic confining potential, which causes the wavepacket to oscillate backwards and forwards
through the avoided crossing:
\[
	V(x) = \alpha x^2 + \begin{pmatrix} \tfrac{1}{2} \tanh(x) & \delta \\ \delta & - \tfrac{1}{2}\tanh(x) \end{pmatrix},
\]
where $X = \delta$, $Z = \tfrac{1}{2}\tanh(x)$, $\rho = \sqrt{\tfrac{1}{4}\tanh(x)^2 + \delta^2}$ and $d(x) = \alpha x^2$.
We choose $\alpha = 0.05$, which gives a relatively weak confining potential.  We use the same grid and time
step as for the simple case in Section~\ref{S:simple} but here evolve back to $t=-5$ and forwards to $t=30$, which gives 3 complete
transitions of the avoided crossing.  Here we start with a wavepacket of the form~\eqref{GaussianP}
on the lower level with $x_0 = 0$ and $p_0 = 5$.  Again we choose $\epsilon = 1/50$. See Figure~\ref{Fig:Potentials} for the adiabatic potentials.

\begin{figure}
\includegraphics[width = \textwidth]{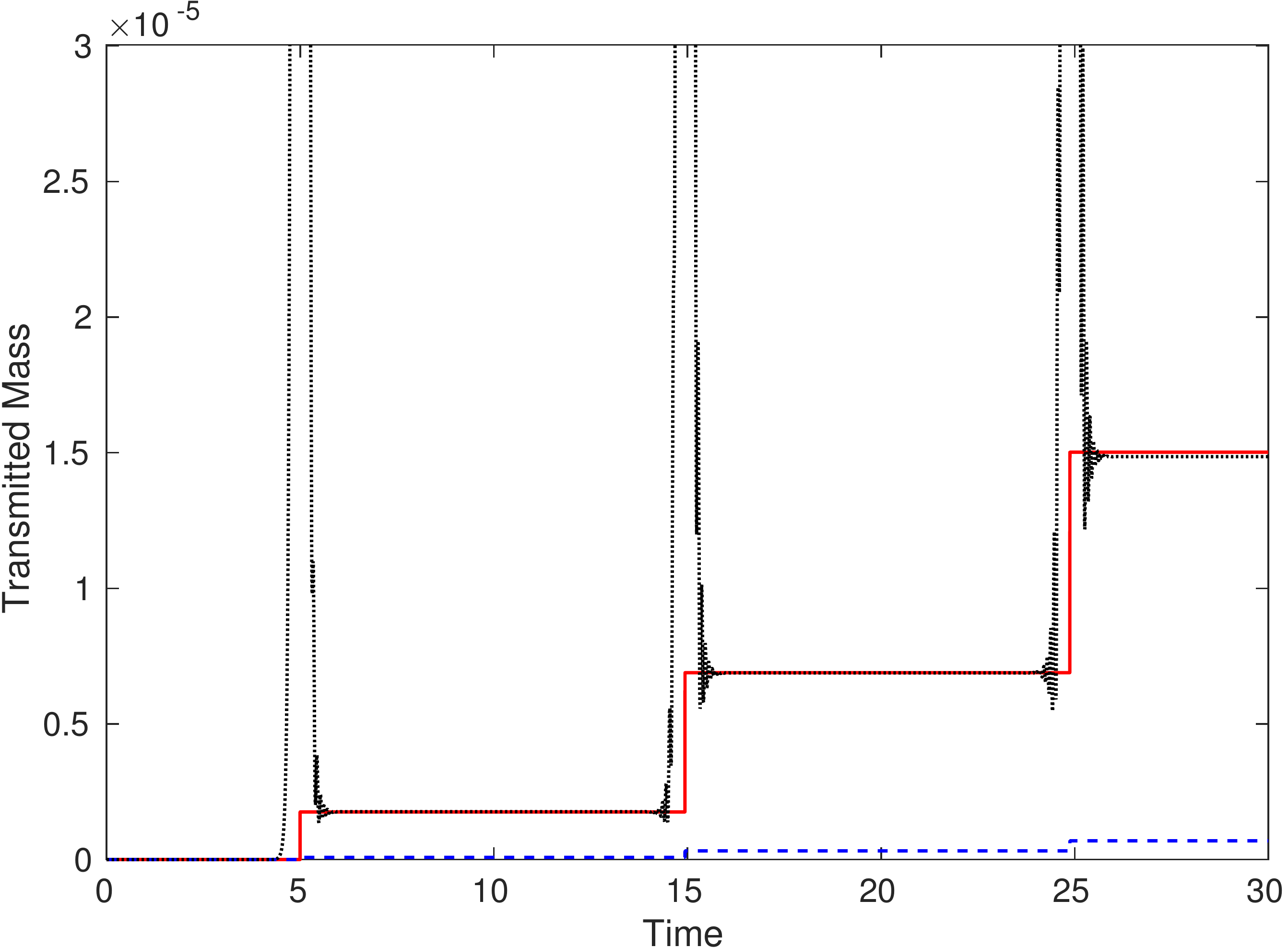}
\caption{Mass of transmitted wavepacket on the upper adiabatic level over time for the `exact' dynamics [black, dotted]
and using the algorithm with formulas~\eqref{formula} [red, solid] and~\eqref{LZformula} [blue, dashed].  
The centre of mass of the wavepacket on the lower level
reaches the avoided crossing three times, at approximately $t = 5, 15, 25$, as indicated by the jumps in the
formula masses.}
\label{Fig:QuadMass}
\end{figure}

As can be seen in Figure~\ref{Fig:QuadMass}, the `exact' dynamics require extreme numerical cancellations 
at each transition in order to produce the true wavepacket.  Although not shown in the Figure, the maximum 
transmitted mass is 0.0028, which is around 200 times larger than the final mass.  Note that the results of 
both formulas are of a similar accuracy to the results for a single crossing, with relative errors 0.0123 and
3.637 for~\eqref{formula} and~\eqref{LZformula}, respectively.  In particular, the agreement between the `exact'
and formula~\eqref{formula} results is excellent whilst, in this case, \eqref{LZformula} significantly underestimates
the size of the transmitted wavepackets.

Whist, in principle, we would expect the results of using~\eqref{LZformula} to improve when $\delta$ decreases
(i.e.\ when the transmitted wavepacket is larger) this adds a complication to the algorithm:  When the transmitted
wavepacket is large, the transition significantly affects the wavepacket on the original level, which is used explicitly in the 
formula for the transmitted wavepacket at the next avoided crossing.  
Due to the perturbative nature of the derivation of the formula \eqref{formula} 
(see e.g.\ \cite{BetzGoddardTeufeul09,BetzGoddard09}), 
the wavepacket on the original level is not treated explicitly, and so we do not have access to this unless
it can be assumed that it is largely unaffected by the transition. 
A necessary requirement for this, due to mass conservation, is that the transmitted wavepacket is small.

\subsection{Dual Avoided Crossings}

\begin{figure}
\includegraphics[width = \textwidth]{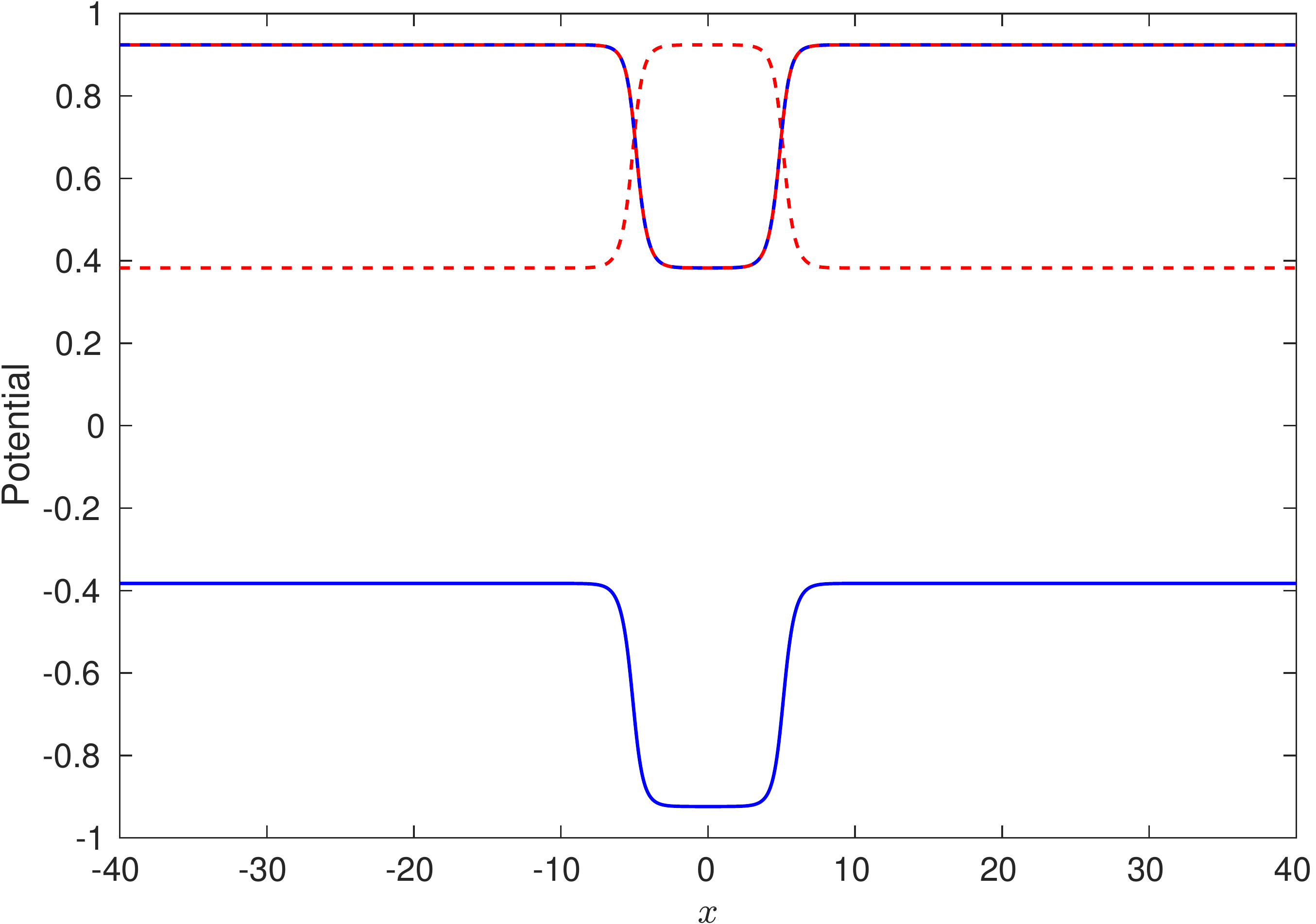}
\caption{Components of the two diabatic eigenvectors $\Phi_1$ and $\Phi_2$ in red and blue, respectively.  
Solid and dashed lines show the two components, denoted by $\Phi_j^\pm$.
Note that $\Phi_1^+ = \Phi_2^-$ and $\Phi_2^+ = -\Phi_1^-$.  The avoided crossings are at $x \approx \pm 5$.}
\label{Fig:DualPotentials}
\end{figure}

As we have seen, for multiple transitions at a single avoided crossing, the algorithm described in Section~\ref{S:algorithm}
works as expected, determining the correct phase of the wavepackets, and therefore also the correct interference
effects.  However, for transitions at separate avoided crossings there is an extra difficulty that arises from the definition
of the diabatic and adiabatic potentials.  As can be seen from Figure~\ref{Fig:DualPotentials}, in an example with two
identical avoided crossings at $\pm x_c$, 
the diabatic eigenfunctions are even. Hence, treating the two crossings independently, the dynamics through the second crossing 
could be computed by flipping the surfaces in space (which gives the diabatic surfaces associated with the first crossing) 
and reversing the momentum of the wavepacket.
From \eqref{formula}, we see that reversing the momentum introduces a sign change in the transmitted wavepacket, 
which must be taken into account when computing the total transmitted wavepacket.  
(This issue is related to the diabatic eigenfunctions only being defined up to their sign.) Note that this argument generalises
to the case where there are multiple non-identical crossings; the case here was chosen for clarity.

Here we choose potentials
\[
	V(x) = \begin{pmatrix} \tfrac{1}{2} \big( \tanh(x - 5) + \tanh(x + 5) + 1 \big) & \delta \\ \delta 
	& - \tfrac{1}{2} \big( \tanh(x - 5) + \tanh(x + 5) + 1 \big) \end{pmatrix}
\]
where we have $X = \delta$ and $Z = \tfrac{1}{2} \big( \tanh(x - 5) + \tanh(x + 5) + 1 \big)$.
We choose $\epsilon = 1/50$, $\delta = 1/2$ and an initial Gaussian condition using \eqref{GaussianP} with
$x_0 = 0$ and $p_0= 5$.  We use the same numerical scheme as in Section~\ref{S:simple}], first evolving backwards for
$t=5$ and then forwards for $t=10$.
See Figure~\ref{Fig:Potentials} for the adiabatic potentials.

\begin{figure}
\includegraphics[width = \textwidth]{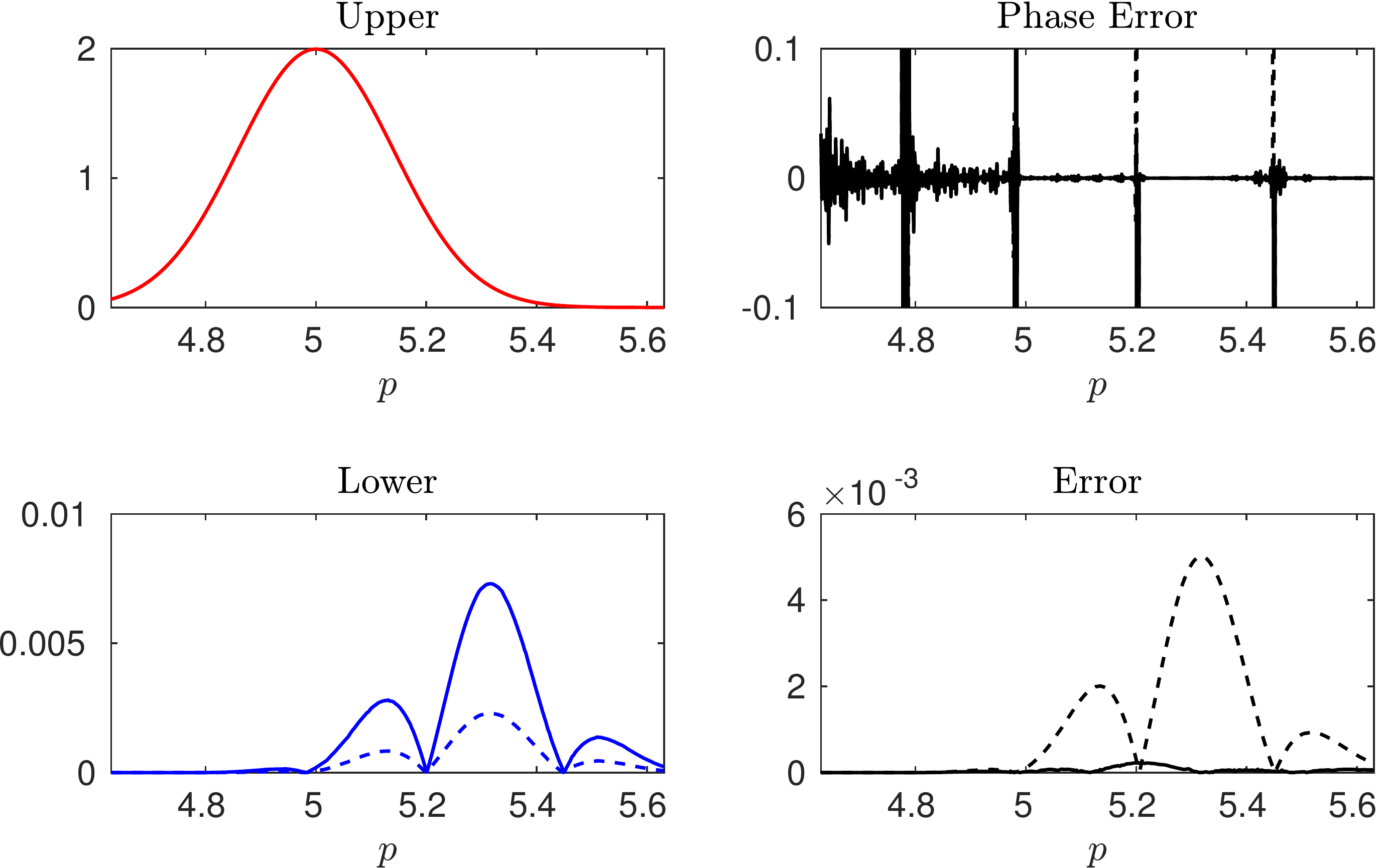}
\caption{Top left: The final wavepacket on the upper level.  Bottom left: The final transmitted wavepackets
on the lower level using \eqref{formula} [solid] and \eqref{LZformula} [dashed].  Bottom right: The associated
errors when compared to the `exact' numerical solution.  Top right: The phase error, which is very small in
both cases apart from when the amplitude of the wavepacket is very small.}
\label{Fig:DualResults}
\end{figure}

As can be seen from Figure~\ref{Fig:DualResults}, the results using~\eqref{formula} are once again very good
(with a relative error of 0.0295), whilst those using the approximate formula~\eqref{LZformula} are much poorer
(relative error 2.193).  Note that if we do not include the additional phase correction described above then the results
using~\eqref{formula} are also very poor.

\bigskip

\section{Conclusions and Open Problems}
\label{S:conclusions}

We have presented a general scheme for the computation of wavepackets transmitted during multiple 
transitions through avoided crossings (at least when the transmitted wavepacket is small), which is also applicable to 
single transitions through full crossings.  In fact, since, in the latter case, almost the entire wavepacket is transmitted,
the scheme should also give accurate results for multiple transitions of full crossings, or combinations of a single
full and multiple avoided crossings.

The principal advantage of our algorithm is that it produces the full quantum wavepacket, including its phase, in
particular allowing the investigation interference effects during multiple transitions.  This is in contrast to standard
surface-hopping algorithms that lose all phase information, and cannot hope to treat systems with interference
effects.

Open problems, which will be the subject of future works, are (i) Approximation of the wavepacket that remains on the original level
when the transmitted wavepacket is not small, which would allow the study of multiple transitions of general
crossings; (ii) Extension to higher dimensions.  This can be done via a slicing algorithm; preliminary results for
model systems are presented in~\cite{BGH18}; (iii) Implementation of our more accurate transition rate in 
surface hopping models, which should extend their range of validity to systems when the transmitted wavepacket is
significantly smaller than those which can be accurately captured by existing LZ schemes.

\acknowledgements
T. Hurst was supported by The Maxwell Institute Graduate School in Analysis and its Applications, a Centre for Doctoral Training funded by the UK Engineering and Physical Sciences Research Council (grant EP/L016508/01), the Scottish Funding Council, Heriot-Watt University and the University of Edinburgh.

\bibliography{LZ}

\end{document}